\documentclass[prb,amsmath,amssymb,superscriptaddress,twocolumn]{revtex4}
\usepackage{graphicx}

\newcommand{\bd}{{\bf d}}
\newcommand{\br}{{\bf r}}

\newcommand{\bk}{{\bf k}}

\newcommand{\bp}{{\bf p}}

\newcommand{\bq}{{\bf q}}

\newcommand{\bG}{{\bf G}}
\newcommand{\bK}{{\bf K}}

\newcommand{\bR}{{\bf R}}

\DeclareMathAlphabet{\mathpzc}{OT1}{pzc}{m}{it} 
\begin{document}
\title{Interacting fermions on the honeycomb bilayer: from weak to strong coupling}
\author{Oskar  Vafek}
\affiliation{National High Magnetic Field Laboratory and Department
of Physics,\\ Florida State University, Tallahasse, Florida 32306,
USA}

\date{\today}
\begin{abstract}
Many-body instabilities of the half-filled honeycomb bilayer are
studied using weak coupling renormalization group as well as strong
coupling expansion. For spinless fermions and assuming parabolic
degeneracy, there are 4-independent four-fermion contact couplings.
While the dominant instability depends on the microscopic values of
the couplings, the broken symmetry state is typically a gapped
insulator with either broken inversion symmetry or broken time
reversal symmetry, with a quantized anomalous Hall effect. Under
certain conditions, the dominant instability may appear in the
particle-particle (pairing) channel. For some non-generic fine-tuned
initial conditions, weak coupling RG trajectories flow into the
non-interacting fixed point, although generally we find runaway
flows which we associate with ordering tendencies. Additionally, a
tight binding model with nearest neighbor hopping and nearest
neighbor repulsion is studied in weak and strong couplings and in
each regime a gapped phase with inversion symmetry breaking is
found. In the strong coupling limit, the ground state wavefunction
is constructed for vanishing in-plane hopping but finite inter-plane
hopping, which explicitly displays the broken inversion symmetry and
a finite difference between the number of particles on the two
layers. Finally, we discuss the spin-1/2 case and use Fierz
identities to show that the number of independent 4-fermion contact
couplings is 9. The corresponding RG equations in the spin-1/2 case
are also presented, and used to show that, just as in strong
coupling, the most dominant weak coupling instability of the
repulsive Hubbard model (at half-filling) is an anti-ferromagnet.
\end{abstract} \maketitle
\section{Introduction}

The problem of interacting fermions on the $A-B$ stacked honeycomb
bilayer at half-filling has attracted attention due to a confluence
of several factors. First, purely on theoretical grounds, in its
simplest form with the nearest neighbor hoping only, the
tight-binding approximation gives rise to a band structure with two
bands touching {\it quadratically} at the Fermi
level\cite{McCannFalko2006PhysRevLett.96.086805,CastroNeto2009RevModPhys.81.109}
near two non-equivalent points in the Brillouin zone, $\bK$ and
$\bK'$. Even at the non-interacting level, such quadratic degeneracy
gives rise to logarithmically divergent
susceptibilities\cite{CastroNeto2009RevModPhys.81.109,
VafekYang2010PhysRevB.81.041401} in several channels as temperature,
or frequency, are taken to
zero\cite{FanZhangPRB2010PhysRevB.81.041402,LevitovPRL2010PhysRevLett.104.156803,Lemonik2010arXiv1006.1399L}.
As a result, some form of spontaneous symmetry breaking is expected
at finite temperature upon inclusion of even weak
interactions\cite{Min2008PhysRevB.77.041407,CastroNeto2009RevModPhys.81.109,KaiSun2009PhysRevLett.103.046811,
VafekYang2010PhysRevB.81.041401,FanZhangPRB2010PhysRevB.81.041402,LevitovPRL2010PhysRevLett.104.156803,Lemonik2010arXiv1006.1399L}.
And while fine-tuning is necessary to achieve such band-structure,
in that (with the exception of square checkerboard and Kagome
lattices studied in Ref.\cite{KaiSun2009PhysRevLett.103.046811})
inclusion of trigonal warping
terms\cite{McCannFalko2006PhysRevLett.96.086805} eventually gives
rise to four Dirac fermions at each $\bK$-point, non-interacting
susceptibilities may be sufficiently enhanced that many-body
instabilities appear, albeit at finite coupling strength. In this
sense, the A-B stacked honeycomb bilayer problem is another example
of the observation that there are no {\it generic} weak coupling
particle-hole
instabilities\cite{RaghuKivelsonScalapino2010PhysRevB.81.224505}.
Rather, fine-tuning, in the form of nesting for example, is
necessary to bring the strong coupling physics down to weak
coupling. If we are interested in accessing the symmetry-broking
phases in the particle-hole channel, as we {\it are} in this case,
then fine tuning is a small price to pay for this access, made
available within perturbative RG. Second, the isolation of graphene
bilayers and the experimental ability to perform, for example,
electrical\cite{Geim2006NatPhys,Yacoby2009NatPhys889,PhilipKim2010PhysRevLett.104.066801},
angle resolved photoemission\cite{RotenbergARPES2006Science313},
Raman spectroscopy\cite{Pinczuk2008PhysRevLett.101.136804} or
infra-red\cite{FengWangIR2009Natur.459..820Z} measurements, while
controlling the gate voltage through the neutrality point, gives
rise to the opportunity to test such theoretical expectations in a
reasonably well controlled physical setting. In addition, the
technological promise of this material fuels further need to
understand its electronic structure and with it the many-body
interactions. Finally, the problem of interacting fermions on the
AB-stacked honeycomb bilayer may soon be realized in cold atom
optical lattices, where the theory may also be tested.

The issue of band-structure fine-tuning notwithstanding, the type of
leading instability in a graphene bilayer (with spin $1/2$ fermions)
has been a subject of debate as well. A mean-field approach has been
used to argue for an insulating state with broken inversion
symmetry\cite{Min2008PhysRevB.77.041407}. A similar approach has
also been argued to lead to trivial gapped insulating
phases\cite{LevitovPRL2010PhysRevLett.104.156803} as well as to an
anomalous quantum Hall
phase\cite{NandkishoreQAH2010arXiv1002.1966N}. On the other hand,
the leading weak coupling instability can be analyzed without
resorting to uncontrolled approximations by using weak coupling
renormalization group. This approach was used in
Ref.\cite{VafekYang2010PhysRevB.81.041401} where a nematic phase was
found to be the dominant instability within the model studied. Such
instability was subsequently also argued for in Ref.
\cite{Lemonik2010arXiv1006.1399L}. On the other hand, an inversion
symmetry breaking insulating phase has been claimed in Ref.
\cite{FanZhangPRB2010PhysRevB.81.041402}.

To determine what type of broken symmetry state is preferred in the
case of {\it spinless} fermions, we perform weak coupling RG
analysis by studying the flow of $4$ independent symmetry allowed
short-range interactions. We find that generically, depending on the
initial values of the 4-fermion contact couplings, the system flows
into a gapped phase with either broken inversion symmetry and a
finite difference between the total number of particles on the two
layers, or broken time reversal symmetry. The former state was not
found to be preferred in the model for spin-$1/2$ fermions studied
in Ref.\cite{VafekYang2010PhysRevB.81.041401} (where the nematic
state was found to dominate), but an example of the latter state
corresponded to one of the fixed points found therein. In
particular, for the spinless case studied here, we find that a
gapped state with anomalous (zero B-field) quantum Hall conductivity
$\pm 2\frac{e^2}{h}$ has the most divergent susceptibility for a
range of initial couplings as determined by the (right) sink of the
RG trajectories shown in Fig.(\ref{fig:BilayerRGflow}). While
non-generic, we also specify special conditions under which the
interacting model flows back to the non-interacting fixed point.

In addition, we analyze the specific microscopic model with nearest
neighbor hopping(s) $t$ (and $t_{\perp}$) and nearest neighbor
repulsion $V$ in both the weak coupling RG and in strong coupling.
In both regimes we find the (trivial) insulating phase with broken
inversion symmetry to dominate. As discussed in more detail below,
in weak coupling the RG flow tends to the left sink shown in
Fig.(\ref{fig:BilayerRGflow}), with a susceptibility that dominates
over other broken symmetry states mainly due to subdominant terms.
In strong coupling, we construct a ground state wavefunction for
$V>0$, $V_{\perp}>0$, $t=0$, but $t_{\perp}\neq 0$, which shows
explicitly the broken layer inversion. Since in this model, the same
symmetry appears to be broken in the limit of both weak and strong
coupling, it is reasonable to assume that such a broken symmetry
state appears at any $V,V_{\perp}>0$.

A similar analysis is presented in the spin-$1/2$ case with short
range interactions. For the repulsive Hubbard model, we find that
the most dominant weak coupling instability is towards an
anti-ferromagnetic state. Since the same ordering tendency happens
in the strong coupling, it is reasonable to assume that in this
model, the Neel ordering appears at any $U>0$.

This paper is organized as follows: in Section II we write down the
(non-interacting) bilayer Hamiltonian first in the tight-binding
approximation and then within $\bk\cdot\bp$ perturbation theory. In
Section III we construct the low energy effective theory at the
neutrality point by fine-tuning the trigonal warping terms to zero.
The rest of that section deals with identifying
microscopic-symmetry-allowed 4-fermion contact interaction terms
using the method of Herbut, Juricic and
Roy\cite{HerbutJuricicRoy2009PhysRevB.79.085116} used for the same
purpose in single-layer graphene. Before the reduction due to Fierz
identities, there are $9$ such couplings which further reduce to 4
once Fierz identities are taken into account. The weak coupling RG
is presented in Section IV, along with the flow diagram in the space
of coupling constant ratios and the analysis of the susceptibility
growth. The $t-V$ model with weak and strong coupling limits is
studied in Section V. In Section VI, the spin-$1/2$ case is
revisited. Symmetry is used to construct an eighteen-dimensional
Fierz vector along with the $18\times 18$ Fierz matrix to show that
there are $9$ independent couplings in this case. Their RG equations
are determined and while more general, they are shown to reduce to
the ones studied in Ref.\cite{VafekYang2010PhysRevB.81.041401} under
conditions outlined therein. In Section VII we study the Hubbard
model in weak and strong coupling. Section VIII is devoted to
conclusions. Details of the derivation are presented in the
Appendices.

\section{Bilayer Hamiltonian}
In this section we will define the non-interacting model by using
two different approximation methods. First, the well known tight
binding approximation \cite{CastroNeto2009RevModPhys.81.109} will be
used and then the $\bk\cdot\bp$-method, or equivalently the method
of
invariants\cite{BirPikusBook1974,McCannFalko2006PhysRevLett.96.086805,AleinerKharzeevTsvelik2007PhysRevB.76.195415,Lemonik2010arXiv1006.1399L}.
Both methods lead to the same form of the low energy Hamiltonian and
it is ultimately a question of convenience which one should be
adopted.
\begin{figure}[h]
\begin{center}
\includegraphics[width=0.5\textwidth]{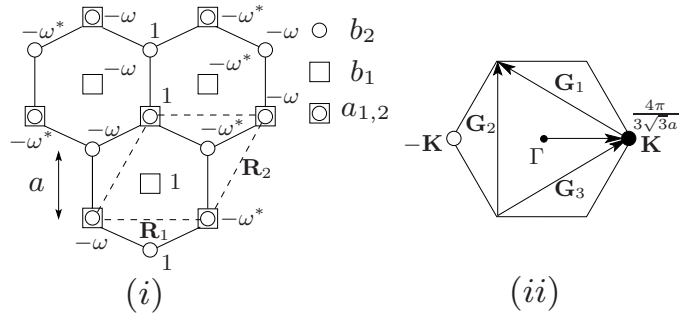}
\end{center}
  \caption{(i) Schematic representation of the A-B stacked bilayer.
  The low energy wavefunction near $\bK$ is also sketched, with
  $\omega=e^{i\pi/3}=\frac{1}{2}+i\frac{\sqrt{3}}{2}$. The primitive lattice vectors
  are $\bR_1=\sqrt{3}a\hat{x}$ and
  $\bR_2=\frac{\sqrt{3}}{2}a\hat{x}+\frac{3}{2}a\hat{y}$. The area
  of the unit cell is $A_{uc}=\hat{z}\cdot(\bR_1\times\bR_2)=\frac{3\sqrt{3}}{2}a^2$.
  (ii) Schematic representation of the (reciprocal) $\bk$-space.
  }\label{fig:BilayerLattice}
\end{figure}

\subsection{Tight-binding approximation}
The non-interacting Hamiltonian in the tight-binding approximation
can be written as
\begin{eqnarray}\label{eq:tightbindingH0}
\mathcal{H}_0&=&H_0^{\perp}+H_0^{\parallel}
\end{eqnarray}
where
\begin{eqnarray}
H_0^{\perp}&=&H_{0,0}^{\perp}+H_{0,1}^{\perp}+H_{0,2}^{\perp}\\
H_{0,0}^{\perp}&=&t_{\perp}\sum_{\bR}\left(a^{\dagger}_{1}(\bR)a_{2}(\bR)+h.c.\right)\\
H_{0,1}^{\perp}&=&t^{(1)}_{\perp}\sum_{\bR,\delta}\left(b^{\dagger}_{1}(\bR+\delta)a_{2}(\bR)
+b^{\dagger}_{2}(\bR-\delta)a_{1}(\bR)+h.c.\right)\nonumber\\\\
H_{0,2}^{\perp}&=&t^{(2)}_{\perp}\sum_{\bR,\delta}\left(
b^{\dagger}_{1}(\bR+\delta_1z)b_{2}(\bR+\delta_1+\delta)+h.c.
\right)\\
H_0^{\parallel}&=&-t\sum_{\bR,\delta}\left(b^{\dagger}_{1}(\bR+\delta)a_{1}(\bR)
+b^{\dagger}_{2}(\bR-\delta)a_{2}(\bR)+h.c.\right)\nonumber\\
\end{eqnarray}
In the case of bilayer graphene, the values of the hopping integrals
$t$, $t_{\perp}$, $t^{(1,2)}_{\perp}$ were extracted experimentally
in Ref.\cite{ZhangBasovFogler2008PhysRevB.78.235408}. If we define
the Fourier transform of a Fermi field as
$c_{j}(\br)=N^{-1/2}_{uc}\sum_{\bk}e^{i\bk\cdot\br}c_{j,\bk}$ where
$c=a$ or $b$, $j=1$ or $2$, and $N_{uc}$ is the number of unit
cells. Next, we let $\chi^{\dagger}_{\bk}=\left(a^{\dagger}_{1,\bk},
a^{\dagger}_{2,\bk}, b^{\dagger}_{2,\bk},b^{\dagger}_{1,\bk}\right)$
to write the non-interacting Hamiltonian (\ref{eq:tightbindingH0})
as
\begin{eqnarray}
\mathcal{H}_0=\sum_{\bk}\chi^{\dagger}_{\bk}\left(\begin{array}{cccc} t_{\perp} & 0 & t^{(1)}_{\perp}d^*_{\bk} & -td_{\bk}\\
0 & t_{\perp} &  -td^*_{\bk} &  t^{(1)}_{\perp}d_{\bk}\\
t^{(1)}_{\perp}d_{\bk} & -td_{\bk}& 0 & t^{(2)}_{\perp}d^*_{\bk}\\
-td^*_{\bk} & t^{(1)}_{\perp}d^*_{\bk} & t^{(2)}_{\perp}d_{\bk}& 0
\end{array}\right)
\chi_{\bk}\end{eqnarray} In the above, the wavevector dependent
function $d_{\bk}=\sum_{\delta}e^{i\bk\cdot\delta}$ where the sum
runs over $\delta_1=\frac{\sqrt{3}}{2}\hat{x}a+\frac{1}{2}\hat{y}a$,
$\delta_2=-\frac{\sqrt{3}}{2}\hat{x}a+\frac{1}{2}\hat{y}a$ and
$\delta_3=-\hat{y}a$. Near $\bK$, $d_{\bK+\bk}\approx
-\frac{3}{2a}t(k_x+ik_y)=-v_Fk_+$. Near $-\bK$, $d_{-\bK+\bk}\approx
\frac{3}{2a}t(k_x-ik_y)=v_Fk_-$. The low energy spectrum of this
(well-known)
Hamiltonian\cite{McCannFalko2006PhysRevLett.96.086805,nilsson:045405},
which is easily diagonalized, will be discussed in the next section.

\subsection{$\bk\cdot\bp$ approach}
Instead of resorting to the tight-binding approximation, we can also
arrive at the low energy Hamiltonian by analyzing the symmetry of
the bilayer potential alone. This is a standard technique when
dealing with semiconductors\cite{BirPikusBook1974} and one which has
also been applied to
graphene\cite{AleinerKharzeevTsvelik2007PhysRevB.76.195415}. For the
sake of self-inclusiveness, we present this method as well to show
that one arrives at the same general form of the Hamiltonian as in
the tight-binding approximation, although in practice the
coefficients of various symmetry-allowed terms must be determined
from experiment. We start with the Schrodinger equation for a
particle moving in potential due to the atoms in layers $1$ and $2$
separated by $2c$
\begin{eqnarray}
\mathcal{H}_0=\frac{\bp^2}{2m_e}+\frac{p^2_z}{2m_e}+V_1(\br)+V_2(\br).
\end{eqnarray}
where
\begin{eqnarray}
V_1(\br)\!&=&\!\sum_{\bR}\left(V_0(\br-\bR-c\hat{z})+V_0(\br-\bR-\delta_1-c\hat{z})\right)\\
V_2(\br)\!&=&\!\sum_{\bR}\left(V_0(\br-\bR+c\hat{z})+V_0(\br-\bR+\delta_1+c\hat{z})\right)
\end{eqnarray}
The low energy field theory is written in terms of the
eight-component Fermi fields (two layers, $1$ and $2$, two valleys,
$\bK$ and $-\bK$, and two sublattices $a$ and $b$ as sketched in
Fig.(\ref{fig:BilayerLattice})):
\begin{eqnarray}
\Psi(\br)&=&\sum_{j=1,2}\left(u_{\bK}^{(a_j)}(\br)\psi_{\bK}^{(a_j)}(\br)+u_{\bK}^{(b_j)}(\br)\psi_{\bK}^{(b_j)}(\br)\right.\nonumber\\
&+&\left.
u_{-\bK}^{(a_j)}(\br)\psi_{-\bK}^{(a_j)}(\br)+u_{-\bK}^{(b_j)}(\br)\psi_{-\bK}^{(b_j)}(\br)\right).
\end{eqnarray}
The rapidly-varying Bloch functions at $\bK$ and at $\bK'=-\bK$ are
related by complex conjugation, $u_{\bK}(\br)=u^*_{-\bK}(\br)$,
irrespective of the layer or sublattice index. Moreover, the Bloch
functions $u^{a_j}_{\bK}(\br)$ and $u^{b_j}_{\bK}(\br)$ transform
irreducibly under point group operations of the lattice (see
Fig.\ref{fig:BilayerLattice}). For the sake of concreteness, within
the nearly free electron approximation for electron wavefunctions
$|\chi_{1,2}\rangle$ confined to layers $1$ and $2$ respectively we
have
\begin{eqnarray}
|u^{(a_1)}_{\bK}\rangle &=&\frac{|\chi_1\rangle}{\sqrt{3}}\left(|\bK\rangle+|\bK+\bG_1\rangle+|\bK-\bG_3\rangle\right)\\
|u^{(b_1)}_{\bK}\rangle &=&\frac{|\chi_1\rangle}{\sqrt{3}}\left(|\bK\rangle-\omega^*|\bK+\bG_1\rangle-\omega|\bK-\bG_3\rangle\right)\\
|u^{(a_2)}_{\bK}\rangle &=&\frac{|\chi_2\rangle}{\sqrt{3}}\left(|\bK\rangle+|\bK+\bG_1\rangle+|\bK-\bG_3\rangle\right)\\
|u^{(b_2)}_{\bK}\rangle
&=&\frac{|\chi_2\rangle}{\sqrt{3}}\left(|\bK\rangle-\omega|\bK+\bG_1\rangle-\omega^*|\bK-\bG_3\rangle\right)
\end{eqnarray}
where $\omega=e^{i\pi/3}=\frac{1}{2}+i\frac{\sqrt{3}}{2}$.
\begin{eqnarray}
\langle u^{(a_j)}_{\bK}|\mathcal{H}_0|u^{(a_j)}_{\bK}\rangle &=&
\langle u^{(b_j)}_{\bK}|\mathcal{H}_0|u^{(b_j)}_{\bK}\rangle=
E_0\nonumber\\
 \langle u^{(a_2)}_{\bK}|\mathcal{H}_0|u^{(a_1)}_{\bK}\rangle
&=& E_0+ \langle
u^{(a_2)}_{\bK}|\hat{V}_2|u^{(a_1)}_{\bK}\rangle\equiv
E_0+t_{\perp}\nonumber\\
 \langle u^{(b_i)}_{\bK}|\mathcal{H}_0|u^{(a_j)}_{\bK}\rangle
&=&E_0\end{eqnarray} i.e. the interlayer hopping arises from the
mixing of the sublattices $a_1$ and $a_2$. The matrix elements of
the in-plane momentum operator $\bp$ are also dictated by symmetry
to be
\begin{eqnarray}
\langle u^{(a_j)}_{\bK}|\bp|u^{(a_j)}_{\bK}\rangle &=& \langle
u^{(b_j)}_{\bK}|\bp|u^{(b_j)}_{\bK}\rangle=\langle u^{(a_1)}_{\bK}|\bp|u^{(a_2)}_{\bK}\rangle=0\nonumber\\
\langle u^{(a_2)}_{\bK}|\bp|u^{(b_2)}_{\bK}\rangle&=&\langle
u^{(b_1)}_{\bK}|\bp|u^{(a_1)}_{\bK}\rangle\sim \hat{x}-i\hat{y}\nonumber\\
\langle u^{(a_1)}_{\bK}|\bp|u^{(b_2)}_{\bK}\rangle&=&\langle
u^{(b_1)}_{\bK}|\bp|u^{(a_2)}_{\bK}\rangle\sim \hat{x}-i\hat{y}\nonumber\\
\langle u^{(b_2)}_{\bK}|\bp|u^{(b_1)}_{\bK}\rangle &\sim&
\hat{x}-i\hat{y}.
\end{eqnarray}
Defining
$\xi^{\dagger}_{\bK}(\br)=({\psi^{(a_1)}_{\bK}}^{\dagger}(\br),
{\psi^{(a_2)}_{\bK}}^{\dagger}(\br),
{\psi^{(b_2)}_{\bK}}^{\dagger}(\br),
{\psi^{(b_1)}_{\bK}}^{\dagger}(\br))$, gives us the effective
Hamiltonian near $\bK$ to read
\begin{eqnarray}
\int d^2\br \xi^{\dagger}_{\bK}(\br)\left(\begin{array}{cccc}
0 & t_{\perp} & v_2k_- & v_F k_+\\
t_{\perp} & 0 & v_F k_- & v_2 k_+ \\
v_2 k_+ & v_Fk_+ & 0 &v_1k_- \\
v_Fk_- & v_2k_- & v_1k_+ &0
\end{array}\right)\xi_{\bK}(\br).
\end{eqnarray}
This is equivalent to what we found in the tight-binding
approximation.

The spectra of the $\bk\cdot\bp$ and the tight-binding Hamiltonians
are well known and have been discussed extensively in the literature
(See e.g.
\cite{McCannFalko2006PhysRevLett.96.086805,ZhangBasovFogler2008PhysRevB.78.235408,CastroNeto2009RevModPhys.81.109}).
In the vicinity of each $\bK$-point, there are four Dirac points:
one isotropic at $\pm\bK$ and three anisotropic ones arranged in
accordance with 3-fold lattice symmetry around the isotropic one.
When we neglect trigonal warping terms, by setting $v_1=v_2=0$, or
set the higher order hopping terms
$t^{(1)}_{\perp}=t^{(2)}_{\perp}=0$, the four Dirac points merge
into a parabolic degeneracy.

\section{Low energy effective theory}
In the weak coupling limit, the kinetic energy dictates which modes
are important to determine the behavior of the system at low
energies. Clearly, at $\bk=0$ we have two degenerate levels and two
levels at $\pm t_{\perp}$. Since we wish to work with a theory for
the low energy modes only, we need to project out the bands which
originate from the two "split-off" bands. We can do so in several
equivalent ways. The method used here implements the path integral
formalism, where we integrate out the Fermi fields associated with
$a_1$ and $a_2$ modes (sites), and arrive at an effective action
with an effective "Hamiltonian" for the low energy modes. In
addition to the wave vector dependence, this "Hamiltonian" is
frequency dependent as well. Near the $\bK$-point, the effective
quadratic action after integrating out the $a$-modes is
\begin{eqnarray}
&&e^{-S^{(0)}_{eff}}=e^{-\int_0^{\beta}d\tau
{\psi^b}^*\left[\partial_{\tau}+H_{bb}\right]\psi^b}\nonumber\\
&\times&\int\mathcal{D}\left({\psi^a}^*\psi^a\right)e^{-\int_0^{\beta}d\tau
\left({\psi^a}^*\left[\partial_{\tau}+H_{aa}\right]\psi^a+{\psi^a}^*H_{ab}\psi^b+{\psi^b}^*H_{ba}\psi^a\right)}.\nonumber
\end{eqnarray}
Since the integral is Gaussian, we can easily perform it and find
that up to an additive constant $S^{(0)}_{eff}=$
\begin{eqnarray}
\frac{1}{\beta}\sum_n{\psi^b}^*(i\omega_n)\left[-i\omega_n+H_{bb}-H_{ab}G_{aa}(i\omega_n)H_{ba}\right]\psi^b(i\omega_n)\nonumber
\end{eqnarray}
where
\begin{equation}\label{eq:Heff1}
H_{ab}G_{aa}(i\omega_n)H_{ba}=\frac{1}{t^2_{\perp}+\omega^2_n}\left[\begin{array}{cc}
\mathcal{A}\bk^2 & \mathcal{B}k^2_+\\
\mathcal{B}k^2_- & \mathcal{A}\bk^2
\end{array}\right]
\end{equation}
\begin{eqnarray}
\mathcal{A}&=&\left(i\omega_n(v^2_F+v^2_2)+2t_{\perp}v_Fv_2\right),\\
\mathcal{B}&=&\left(t_{\perp}(v^2_F+v^2_2)+2i\omega_nv_Fv_2\right),
\end{eqnarray}
and
\begin{equation}\label{eq:Heff2}
H_{bb}=\left[\begin{array}{cc}
0 & v_1k_-\\
v_1k_+ & 0
\end{array}
 \right].
\end{equation}
Within the $\bk\cdot\bp$ theory, the parameters $v_1$ and $v_2$
should be determined from experiment. To make contact with the
notation in literature, Ref.\cite{nilsson:045405} have $v_1=v_3$ and
$v_2=-v_4$ (see their Eqs. 6 and 15).

If we are interested in the modes near the Fermi level of an
unbiased bilayer, we can simply set $\omega_n=0$ in the effective
action (\ref{eq:Heff1}-\ref{eq:Heff2}). As will be obvious from the
discussion in the next section, terms arising from the corrections
are perturbatively irrelevant near the Gaussian fixed point in the
sense discussed in a different context in
Ref.\cite{Shankar.RevModPhys.66.129.1994}.

In what follows we will also set $v_1=v_2=0$ to fine-tune the system
to quadratic degeneracy. Such a situation arises if in the
tight-binding formulation we consider only the nearest neighbor
hopping integrals, $t$ and $t_{\perp}$. Otherwise, as mentioned in
the introduction, the ultimate low energy dispersion involves four
(one isotropic and three anisotropic) Dirac
cones\cite{McCannFalko2006PhysRevLett.96.086805,nilsson:045405,Lemonik2010arXiv1006.1399L}.
While such fine-tuning appears artificial, it is an example of the
maxim\cite{RaghuKivelsonScalapino2010PhysRevB.81.224505} that there
are no {\it generic} weak coupling particle-hole instabilities.
Rather, fine-tuning, in the form of nesting for example, is
necessary to bring the strong coupling physics down to weak
coupling. If we are interested in accessing the symmetry-breaking
phases in the particle-hole channel, as we {\it are} in this case,
then fine tuning is a small price to pay for this access made
available within perturbative
RG\cite{RaghuKivelsonScalapino2010PhysRevB.81.224505}.

Putting back the $-\bK$ point, the low energy degrees of freedom can
now be expressed in terms of a four component Fermi field
$$\psi^{\dagger}(\br)=\left({\psi_{\bK}^{(b_1)}}^{\dagger}(\br),{\psi_{\bK}^{(b_2)}}^{\dagger}(\br),{\psi_{-\bK}^{(b_1)}}^{\dagger}(\br),
{\psi_{-\bK}^{(b_2)}}^{\dagger}(\br)\right),$$ i.e. the electronic
degrees of freedom are expanded as
\begin{eqnarray}\label{eq:psiDefinition}
\tilde{\Psi}(\br)&=&\sum_{j=1,2}\left(u_{\bK}^{(b_j)}(\br)\psi_{\bK}^{(b_j)}(\br)+
u_{-\bK}^{(b_j)}(\br)\psi_{-\bK}^{(b_j)}(\br)\right).
\end{eqnarray}

The non-interacting low energy (imaginary time $\tau$) Lagrangian,
which includes both $K$ and $K'$ valleys, and which will serve as
our (gaussian) fixed point of departure, can therefore be written as
\begin{eqnarray}\label{eq:Lag0}
\mathcal{L}_0&=& \int d^2\br
\left[\psi^{\dagger}(\tau,\br)\left(\frac{\partial}{\partial
\tau}+\sum_{a=x,y}\Sigma^ad^a_{\bp}\right)\psi(\tau,\br)\right]
\end{eqnarray}
where we defined the vector function $\bd_{\bk}$ and the $4\times4$
matrices $\Sigma^{x,y}$ as
\begin{eqnarray}
d^x_{\bk}&=&\frac{k^2_x-k^2_y}{2m},\;\;\; d^y_{\bk}=-\frac{2k_xk_y}{2m}\\
\Sigma^x&=&1\sigma^x=\gamma_2,\;\;\;\Sigma^y=\tau^z\sigma^y=\gamma_1.
\end{eqnarray}
The effective mass parameter entering the above equations is
$m=t_{\perp}/(2v^2_F)$ (In the tight binding approximation
$v_F=3t/(2a)$). The four component Fermi objects $\psi$ appearing in
Eq. (\ref{eq:Lag0}) were defined as the envelope Fermi fields in
Eq.(\ref{eq:psiDefinition}). In the above, the first Pauli matrix
acts in the valley $\pm\bK$-space and the second in the layer
$1,2$-space. To make contact with the literature we also use Dirac
$\gamma$-matrices which we represent as
\begin{eqnarray}
\gamma_0&=&1\sigma^z\\
\gamma_1&=&\tau^z\sigma^y\\
\gamma_2&=&1\sigma^x\\
\gamma_3&=&\tau^x\sigma^y\\
\gamma_5&=&\tau^y\sigma^y.
\end{eqnarray}

The action $\int d\tau \mathcal{L}_0$ is invariant under the scale
transformation
\begin{eqnarray}\label{eq:gaussianScaling}
\br &\rightarrow& s\br\\
\tau &\rightarrow& s^2\tau\\
\psi &\rightarrow& s^{-1}\psi.
\end{eqnarray}
This means that the "dynamical critical exponent" $z=2$ for the
gaussian theory, which will be our point of departure when analyzing
{\it weak coupling} instabilities.

\subsection{Short range interactions}
From the above discussion of the gaussian fixed point, it is evident
that the short range interactions, when projected onto our low
energy modes, will contain among other (perturbatively irrelevant)
terms, contact four-fermion terms which are marginal by power
counting. The rest of this section deals with identifying such
symmetry-allowed interaction terms. The method used here follows
almost verbatim the method used by Herbut, Juricic and
Roy\cite{HerbutJuricicRoy2009PhysRevB.79.085116} in their analysis
of the short range interactions in {\it single} layer graphene. In
addition to the lattice symmetries used in
Ref.\cite{HerbutJuricicRoy2009PhysRevB.79.085116}, we also include
the three-fold rotational
symmetry\cite{AleinerKharzeevTsvelik2007PhysRevB.76.195415}, which
reduces the number of independent four-fermion couplings to $4$.

We can therefore start by writing the general Lagrangian
\begin{eqnarray}
\mathcal{L}&=&\mathcal{L}_0+\mathcal{L}_{int},
\end{eqnarray}
where $\mathcal{L}_0$ was introduced in Eq.(\ref{eq:Lag0}) and
\begin{eqnarray}\label{eq:LintBeforeFierz0}
\mathcal{L}_{int}&=& \frac{1}{2}\sum_{S,T}g_{ST}\int d^2\br
\left(\psi^{\dagger}S\psi(\br,\tau)\right)\left(\psi^{\dagger}
T\psi(\br,\tau)\right)
\end{eqnarray}
where, at this point, the sum over $S$ includes all sixteen
independent four-by four matrices (generators of $SU(4)$) and so
does the sum over $T$. Naively, we have $16+8*15=136$ couplings to
consider. Just as in the case of the single-layer
graphene\cite{HerbutJuricicRoy2009PhysRevB.79.085116}, this number
will be dramatically reduced first by using the discrete symmetries
of the lattice and second by using Fierz identities.

The key role in this reduction is played by the behavior of the
Bloch functions $u(\br)$ under symmetry operations, which dictates
the transformation properties of the four component, slowly varying,
envelope Fermi fields
$\psi(\br)$\cite{BirPikusBook1974,AleinerKharzeevTsvelik2007PhysRevB.76.195415}.
The dimer centered rotation by $2\pi/3$, mirror reflection about the
yz-plane and about the xz-axis followed by xy-plane respectively
give
\begin{eqnarray}
\hat{C}_3\psi(x,y)&=&-e^{-i\frac{\pi}{3}\tau_3\sigma_3}\psi\left(-\frac{1}{2}x-\frac{\sqrt{3}}{2}y,\frac{\sqrt{3}}{2}x-\frac{1}{2}y\right)\nonumber\\
&=&-e^{\frac{\pi}{3}\gamma_1\gamma_2}\psi\left(-\frac{1}{2}x-\frac{\sqrt{3}}{2}y,\frac{\sqrt{3}}{2}x-\frac{1}{2}y\right)\label{eq:C3rotations}\\
\hat{\sigma}^y_v\psi(x,y)&=&\tau_1 1_2\psi(-x,y)=i\gamma_1\gamma_{5}\psi(-x,y)\label{eq:reflectionY}\\
\hat{\sigma}^z_v\hat{\sigma}^x_v\psi(x,y)&=&
1_2\sigma_1\psi(x,-y)=\gamma_2\psi(x,-y)\label{eq:reflectionX}
\end{eqnarray}
The time reversal symmetry and translational symmetry give
\begin{eqnarray}
\Theta\psi(\br)&=&\tau_11_2\psi^*(\br)=i\gamma_1\gamma_5\psi^*(\br)\label{eq:timeReversal}\\
\hat{t}\psi(\br)&=&e^{i\bK\cdot\bR\tau_31_2}\psi(\br+\bR)=e^{\bK\cdot\bR
\gamma_3\gamma_5}\psi(\br+\bR).\label{eq:translation}
\end{eqnarray}
In the above, $\bR=m\bR_1+n\bR_2$ where $\bR_1=\sqrt{3}a\hat{x}$ and
$\bR_2=\frac{\sqrt{3}}{2}a\hat{x}+\frac{3}{2}a\hat{y}$. And since
$\bK=\frac{4\pi}{3\sqrt{3}a}\hat{x}$,
$\bK\cdot\bR=\frac{2\pi}{3}(2m+n)$, where $m,n=0,\pm1,\pm2,\ldots$.
The lattice translational symmetry therefore corresponds to the
$Z_3$ discrete analog of the chiral $U_c(1)$ generated by
$\gamma_3\gamma_5$.

\subsubsection{Symmetry reduction}

Following Herbut {\it
et.al.}\cite{HerbutJuricicRoy2009PhysRevB.79.085116}, we split the
sixteen linearly independent four-by-four matrices $S$ and $T$ into
four sets
\begin{eqnarray}\label{eq:su4generators}
A&=&\{1_4,\gamma_2,i\gamma_0\gamma_3,i\gamma_1\gamma_5\}\\
B&=&\{i\gamma_0\gamma_1,-i\gamma_3\gamma_5,i\gamma_0\gamma_5,i\gamma_1\gamma_3\}\\
C&=&\{\gamma_0,i\gamma_0\gamma_2,\gamma_3,i\gamma_2\gamma_3\}\\
D&=&\{\gamma_1,i\gamma_1\gamma_2,\gamma_5,i\gamma_2\gamma_5\}.
\end{eqnarray}
The matrices which belong to the set $A$ are even under both
reflection operations (\ref{eq:reflectionY}) and
(\ref{eq:reflectionX}). The matrices in the set $B$ are odd under
$y$-reflections (\ref{eq:reflectionY}) and even under
"$x$"-reflections (\ref{eq:reflectionX}). The matrices in the set
$C$ are even under $y$-reflections (\ref{eq:reflectionY}) and odd
under "$x$"-reflections (\ref{eq:reflectionX}). And finally,
matrices belonging to the set $D$ are odd under both
(\ref{eq:reflectionY}) and (\ref{eq:reflectionX}). This means that
only quartic terms combining matrices from the same set are allowed
by symmetry. Each such set contains $4+2*3=10$ such terms and that
leaves $40$ couplings.

Eight matrices $A_{1,2}$, $B_{1,2}$, $C_{1,2}$ and $D_{1,2}$ are
left invariant under the spatial translation operation
(\ref{eq:translation}). These give rise to $3*4=12$ couplings, eight
direct $g_{X_jX_j}$ ($X=A,B,C,\mbox{or} D$ and $j=1,\mbox{or} 2$),
as well as four mixed $g_{X_1X_2}$. In addition, there are four sets
of pairs which transform as vectors under (\ref{eq:translation}):
$\alpha=\{A_{3},B_{3}\}$, $\beta=\{B_{4},A_{4}\}$,
$\gamma=\{C_3,D_3\}$ and $\delta=\{C_4,D_4\}$. These give rise to
additional $6$ couplings. Schematically, four of them are
$\sum_{\rho=\alpha,\beta,\gamma,\delta}\sum_{j=1}^2g_{\rho}\rho_j\otimes\rho_j$
and two mixed ones are $g_{\alpha\beta}(A_3\otimes A_4-B_3\otimes
B_4)$ and $g_{\gamma\delta}(C_3\otimes C_4+D_3\otimes D_4)$.
Altogether, after inclusion of the translation symmetry, we are left
with $18$ couplings.

The unitary part of the time reversal operations $\Theta$, Eq.
(\ref{eq:timeReversal}), happens to correspond to the mirror
reflection about $y$, (Eq.\ref{eq:reflectionY}), which has already
been taken into account. However, complex conjugation, further
restricts the number of couplings. Specifically, mixed terms with
one purely real and one purely imaginary matrix cannot appear,
therefore $g_{C_1C_2}=g_{D_1D_2}=g_{\gamma\delta}=0$. This leaves
$15$ couplings\cite{HerbutJuricicRoy2009PhysRevB.79.085116}.

The lattice symmetries considered by Herbut {\it
et.al}\cite{HerbutJuricicRoy2009PhysRevB.79.085116} did not contain
site- or plaquette- centered
rotation\cite{AleinerKharzeevTsvelik2007PhysRevB.76.195415} by
$120^{o}$. As stated in Eq.(\ref{eq:C3rotations}), this symmetry is
generated by $i\gamma_1\gamma_2$. Including this symmetry requires
that the cross-terms $g_{A_1A_2}=g_{B_1B_2}=g_{\alpha\beta}=0$.
Moreover, it requires that $g_{A_2A_2}=g_{D_1D_1}$,
$g_{B_1B_1}=g_{C_2C_2}$, and $g_{\beta}=g_{\delta}$.

This leaves us with the following 9 terms
\begin{eqnarray}\label{eq:LintBeforeFierz}
&&g_{A_1A_1}(\psi^{\dagger}A_1\psi)^2+g_{A_2A_2}\left[(\psi^{\dagger}A_2\psi)^2+(\psi^{\dagger}D_1\psi)^2\right]\nonumber\\
&+&g_{B_1B_1}\left[(\psi^{\dagger}B_1\psi)^2+(\psi^{\dagger}C_2\psi)^2\right]+g_{B_2B_2}(\psi^{\dagger}B_2\psi)^2\nonumber\\
&+& g_{C_1C_1}(\psi^{\dagger}C_1\psi)^2
+g_{D_2D_2}(\psi^{\dagger}D_2\psi)^2\nonumber\\
&+&
g_{\alpha}\left[(\psi^{\dagger}A_3\psi)^2+(\psi^{\dagger}B_3\psi)^2\right]
+g_{\gamma}\left[(\psi^{\dagger}C_3\psi)^2+(\psi^{\dagger}D_3\psi)^2\right]
\nonumber\\
&+&
g_{\beta}\left[(\psi^{\dagger}B_4\psi)^2+(\psi^{\dagger}A_4\psi)^2+(\psi^{\dagger}C_4\psi)^2+(\psi^{\dagger}D_4\psi)^2\right].
\end{eqnarray}
The 9 terms can be further reduced to 4 independent ones by using
Fierz identities.

\subsubsection{Fierz identities}
\begin{figure}[t]
\begin{center}
\begin{tabular}{ccccc}
\includegraphics[width=0.1\textwidth]{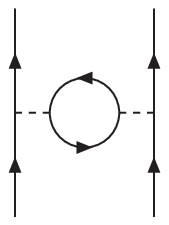}&
\includegraphics[width=0.09\textwidth]{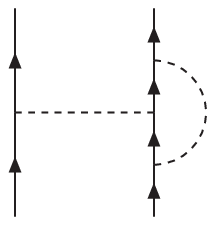}&
\includegraphics[width=0.09\textwidth]{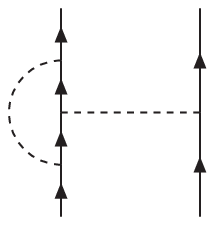}&
\includegraphics[width=0.07\textwidth]{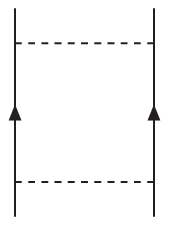}&
\includegraphics[width=0.07\textwidth]{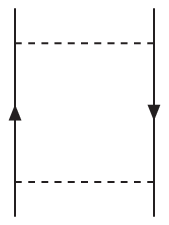}
\end{tabular}
\end{center}
  \caption{Diagrams appearing at 1-loop RG. Each vertex can be
  represented by either a 4$\times$4 matrix (spinless) or 8$\times$8
  matrix (spin-$1/2$) case.
  }\label{fig:graphs}
\end{figure}

We set $g_{XX}=g_{X}$ to continue with the notation of
Ref.\cite{HerbutJuricicRoy2009PhysRevB.79.085116}. We use the method
employed therein to write down Fierz
identities\cite{HerbutJuricicRoy2009PhysRevB.79.085116,ItzyksonZuberBook}
which, due to the Grassman nature of the Fermi fields, relate
various seemingly unrelated couplings.

The starting point is the $SU(4)$ algebraic identity (see Eq.(A4) of
Ref.\cite{HerbutJuricicRoy2009PhysRevB.79.085116})
\begin{eqnarray}
S_{ij}T_{mn}
&=&\frac{1}{16}\mbox{Tr}[S\Gamma^aT\Gamma^b]\Gamma^{b}_{in}\Gamma^a_{mj}
\end{eqnarray}
which leads to
\begin{eqnarray}\label{eq:Fierz4by4start}
&&\left(\psi^{\dagger}(x)S\psi(x)\right)\left(\psi^{\dagger}(y)T\psi(y)\right)=\nonumber\\
&&-\frac{1}{16}\mbox{Tr}[S\Gamma^aT\Gamma^b]\left(\psi^{\dagger}(x)\Gamma^b\psi(y)\right)\left(\psi^{\dagger}(y)\Gamma^a\psi(x)\right).
\end{eqnarray}
The minus sign comes from $\psi$ and $\psi^{\dagger}$ being
anti-commuting (four component) Grassman fields. For contact terms
$x=y$ and the above equation (\ref{eq:Fierz4by4start}) constitutes a
set of linear relations between different terms of our symmetry
reduced interaction Lagrangian (\ref{eq:LintBeforeFierz}).

If we arrange the quartic terms into a vector
\begin{eqnarray}\label{eq:FierzVector}
V&=&\left\{(\psi^{\dagger}A_1\psi)^2,(\psi^{\dagger}A_2\psi)^2+(\psi^{\dagger}D_1\psi)^2,\right.\nonumber\\
&&\left.(\psi^{\dagger}B_1\psi)^2+(\psi^{\dagger}C_2\psi)^2,
(\psi^{\dagger}B_2\psi)^2,(\psi^{\dagger}C_1\psi)^2,\right.\nonumber\\
&&\left.(\psi^{\dagger}D_2\psi)^2,
(\psi^{\dagger}A_3\psi)^2+(\psi^{\dagger}B_3\psi)^2,\right.\nonumber\\
&&\left.
(\psi^{\dagger}B_4\psi)^2+(\psi^{\dagger}A_4\psi)^2+(\psi^{\dagger}C_4\psi)^2+(\psi^{\dagger}D_4\psi)^2,\right.\nonumber\\
&&\left.(\psi^{\dagger}C_3\psi)^2+(\psi^{\dagger}D_3\psi)^2\right\},
\end{eqnarray}
then the Fierz identities lead to the linear constraint
\begin{eqnarray}
FV&=&0.
\end{eqnarray}
A straightforward, though somewhat laborious, application of
(\ref{eq:Fierz4by4start}) leads to the explicit form of the Fierz
matrix in the case of spinless fermions
\begin{eqnarray} F&=&\left(
\begin{array}{ccccccccc}
5& 1& 1& 1& 1& 1& 1& 1& 1\\
2& 4& 0& 2& -2& -2& 2& 0& -2\\
2& 0& 4& 2& -2& -2& -2& 0& 2\\
1& 1& 1& 5& 1& 1& -1& -1& -1\\
1& -1& -1& 1& 5& 1& -1& 1& -1\\
1& -1& -1& 1& 1& 5& 1& -1& 1\\
2& 2& -2& -2& -2& 2& 4& 0& 0\\
4& 0& 0& -4& 4& -4& 0& 4& 0\\
2& -2& 2& -2& -2& 2& 0& 0& 4
\end{array}
\right).
\end{eqnarray}
The matrix $F$ has four zero eigenvalues and as a result there are
four independent
couplings\cite{HerbutJuricicRoy2009PhysRevB.79.085116}.

\begin{widetext}
In order to make a connection with the previous
work\cite{VafekYang2010PhysRevB.81.041401}, we choose to eliminate
\begin{eqnarray}
(\psi^{\dagger}B_1\psi)^2+(\psi^{\dagger}C_2\psi)^2&=&-2(\psi^{\dagger}A_1\psi)^2+\left[(\psi^{\dagger}A_2\psi)^2+(\psi^{\dagger}D_1\psi)^2\right]-2(\psi^{\dagger}D_2\psi)^2
-2\left[(\psi^{\dagger}C_3\psi)^2+(\psi^{\dagger}D_3\psi)^2\right]\nonumber\label{eq:FierzB2}\\\\
(\psi^{\dagger}B_2\psi)^2&=&-\left[(\psi^{\dagger}A_2\psi)^2+(\psi^{\dagger}D_1\psi)^2\right]+(\psi^{\dagger}D_2\psi)^2
+\left[(\psi^{\dagger}C_3\psi)^2+(\psi^{\dagger}D_3\psi)^2\right]\\
\label{eq:FierzC1}
(\psi^{\dagger}C_1\psi)^2&=&-(\psi^{\dagger}A_1\psi)^2+\left[(\psi^{\dagger}A_2\psi)^2+(\psi^{\dagger}D_1\psi)^2\right]-2(\psi^{\dagger}D_2\psi)^2
-\left[(\psi^{\dagger}C_3\psi)^2+(\psi^{\dagger}D_3\psi)^2\right]\\
\label{eq:FierzA3}
(\psi^{\dagger}A_3\psi)^2+(\psi^{\dagger}B_3\psi)^2&=&-2(\psi^{\dagger}A_1\psi)^2-2(\psi^{\dagger}D_2\psi)^2
-\left[(\psi^{\dagger}C_3\psi)^2+(\psi^{\dagger}D_3\psi)^2\right]
\end{eqnarray}
and
\begin{eqnarray}\label{eq:FierzA4}
(\psi^{\dagger}B_4\psi)^2+(\psi^{\dagger}A_4\psi)^2+(\psi^{\dagger}C_4\psi)^2+(\psi^{\dagger}D_4\psi)^2=
-2\left[(\psi^{\dagger}A_2\psi)^2+(\psi^{\dagger}D_1\psi)^2\right]+4(\psi^{\dagger}D_2\psi)^2\nonumber\\
+2\left[(\psi^{\dagger}C_3\psi)^2+(\psi^{\dagger}D_3\psi)^2\right],
\end{eqnarray}
in favor of the remaining four terms. These equations will be used
in deriving our RG equations, since elimination of fast modes will
generate terms such as, for example, $(\psi^{\dagger}B_2\psi)^2$.
The above equations show that such a term does not correspond to a
new coupling in a renormalized action, but rather is a linear
combination of terms already present.

Finally, we arrive at our interaction Lagrangian
\begin{eqnarray}\label{eq:LintAfterFierz}
\mathcal{L}_{int}&=& \frac{1}{2}\int  d^2\br
\left[g_{A_1}\left(\psi^{\dagger}A_1\psi(\br,\tau)\right)^2+
g_{A_2}\left(\left(\psi^{\dagger}A_2\psi(\br,\tau)\right)^2+\left(\psi^{\dagger}D_1\psi(\br,\tau)\right)^2\right)\right]\nonumber\\
&+&\frac{1}{2}\int d^2\br
\left[g_{D_2}\left(\psi^{\dagger}D_2\psi(\br,\tau)\right)^2+
g_{\gamma}\left(\left(\psi^{\dagger}C_3\psi(\br,\tau)\right)^2+\left(\psi^{\dagger}D_3\psi(\br,\tau)\right)^2\right)\right].
\end{eqnarray}
Above is the most general four-fermion contact interaction
Lagrangian for spinless fermions allowed by the symmetry of the A-B
stacked honeycomb bilayer. In the next section, we study the weak
coupling RG flow of the four couplings $g_{A_1},g_{A_2},g_{D_2}$ and
$g_{\gamma}$. The first three couplings appeared in our previous
work\cite{VafekYang2010PhysRevB.81.041401} where we called them
$g_1$, $g_2$, and $g_3$. The fourth coupling, $g_{\gamma}$, did not
appear there since the starting point assumed only finite $g_1$ and,
as we will see later, $g_{\gamma}$ is not generated if its starting
value is zero.
\begin{figure}[t]
\begin{center}
\includegraphics{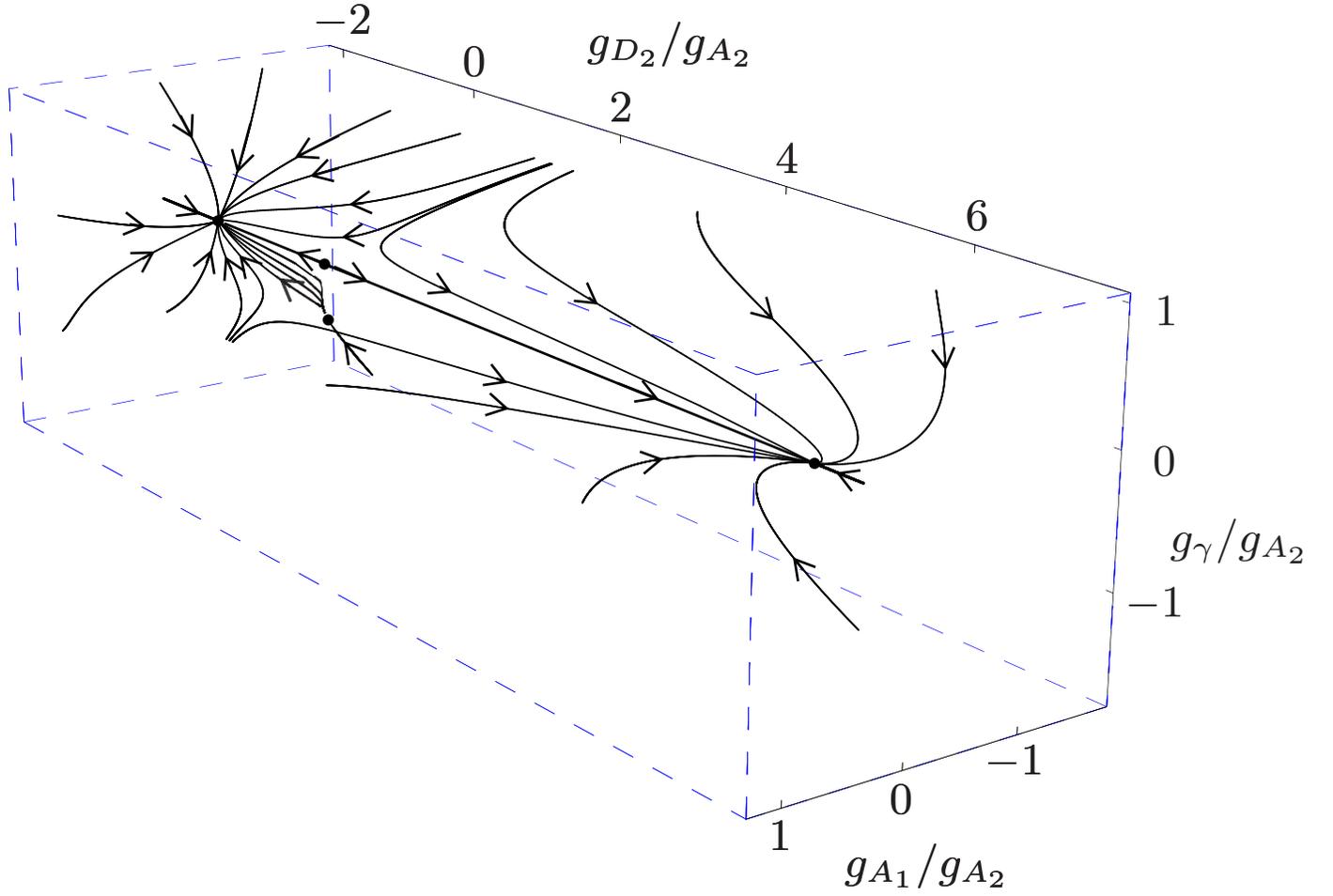}
\end{center}
  \caption{Flow diagram in the coupling constant ratio space
  assuming that $g_{A_2}<0$ (generic behavior). There are two sinks
  given by Eqs.(\protect\ref{eq:fixedRatios1}) and (\protect\ref{eq:fixedRatios3})
  in the text
  and two mixed fixed ratios (\protect\ref{eq:fixedRatios2}) and
  (\protect\ref{eq:fixedRatios4}). For $g_{A_2}>0$ the flow is reversed, and
  generically, the divergent coupling constant ratios simply mean
  that $g_{A_2}$ has shrunk and crossed $0$. After this point
  $g_{A_2}$ becomes negative and the directionality shown here is
  restored.
  }\label{fig:BilayerRGflow}
\end{figure}
\end{widetext}
\section{Renormalization group analysis}
Clearly, $g_{ST}$'s are marginal by power-counting and the question
is how they flow. The RG procedure employed here follows
Ref.\cite{Shankar.RevModPhys.66.129.1994} and consists of
integrating out the fermionic modes in a thin shell between the
initial cutoff $\Lambda$ and $\Lambda/s$, while the integral over
$\omega$ extends from $-\infty$ to $\infty$. Since we are working in
weak coupling, we can integrate out the fast modes perturbatively in
$g$'s. The diagrams needed are shown in Fig.(\ref{fig:graphs}).
Afterwards, the lengths $\br$, times $\tau$ and the modes $\psi$ are
rescaled according to Eqs.(\ref{eq:gaussianScaling}) and the change
of the coupling constants is noted. (To the order we are working,
the dynamical critical exponent $z$ remains $2$). While the details
of the derivation are provided in the Appendix, we note in passing
that the analysis is facilitated by the use of the identities
\begin{eqnarray}\label{eq:greensFxnTrick}
&&\int^{\infty}_{-\infty}\frac{d\omega}{2\pi}\int^{\Lambda}_{\Lambda/s}\frac{d^2\bk}{(2\pi)^2}
G_{\bk}(i\omega)\otimes G_{\mp\bk}(\mp i\omega)=\nonumber\\
&&\left(\pm1_4\otimes1_4+\frac{1}{2}\sum_{a=1}^2\gamma_a\otimes\gamma_a\right)\frac{m}{4\pi}\ln
s
\end{eqnarray}
where the non-interacting Green's function is
\begin{eqnarray}
G_{\bk}(i\omega)&=&\left(-i\omega+\Sigma\cdot
d_{\bk}\right)^{-1}=\frac{i\omega+\Sigma\cdot
d_{\bk}}{\omega^2+\left(\frac{\bk^2}{2m}\right)^2}
\end{eqnarray}
and, just as before, $d^x_{\bk}=\frac{k^2_x-k^2_y}{2m}$,
$d^y_{\bk}=-\frac{2k_xk_y}{2m}$, $\Sigma^x=\gamma_2$, and
$\Sigma^y=\gamma_1$.

Using this procedure, we find the RG equations for the four coupling
constants to be
\begin{eqnarray}\label{eq:RGspinless}
\frac{dg_{A_1}}{d\ln s}&=&-4g_{A_1}g_{A_2}\frac{m}{4\pi}\\
\frac{dg_{A_2}}{d\ln
s}&=&-\left(g^2_{A_1}-2g_{A_1}g_{A_2}+8g^2_{A_2}
-2g_{A_2}g_{D_2}+g^2_{D_2}+\right.\nonumber\\
&&\left.    4\left(3g_{A_2}-g_{D_2}\right)g_{\gamma}+6g^2_{\gamma}\right)\frac{m}{4\pi}\\
\frac{dg_{D_2}}{d\ln
s}&=&2\left(2g^2_{A_2}+2g_{A_1}g_{D_2}-6g_{A_2}g_{D_2}-2g^2_{D_2}+\right.\nonumber\\
&&\left.8g_{A_2}g_{\gamma}+2g^2_{\gamma}\right)\frac{m}{4\pi}\\
\frac{dg_{\gamma}}{d\ln
s}&=&-2g_{\gamma}\left(-2g_{A_1}+2g_{A_2}+2g_{\gamma}\right)\frac{m}{4\pi}.
\end{eqnarray}
These equations reduce to the ones studied in
Ref.\cite{VafekYang2010PhysRevB.81.041401} when we set
$g_{\gamma}=0$ in this work and $N=2$ in Eqs.(6-8) of
Ref.\cite{VafekYang2010PhysRevB.81.041401}. Their analysis proceeds
along the lines discussed in
Ref.\cite{VafekYang2010PhysRevB.81.041401}. We note that each RG
equation corresponds to a quadratic polynomial in coupling
constants. Therefore, dividing each equation by $g_{A_2}$ (which is
$g_3$ in the notation of
Ref.\cite{VafekYang2010PhysRevB.81.041401}), we obtain three
equations
\begin{eqnarray}\label{eq:RGratios00}
\frac{dg_{A_1}}{dg_{A_2}}&=&\mathcal{R}_{12}\left(\frac{g_{A_1}}{g_{A_2}},\frac{g_{D_2}}{g_{A_2}},\frac{g_{\gamma}}{g_{A_2}}\right)\\
\label{eq:RGratios01}
\frac{dg_{D_2}}{dg_{A_2}}&=&\mathcal{R}_{32}\left(\frac{g_{A_1}}{g_{A_2}},\frac{g_{D_2}}{g_{A_2}},\frac{g_{\gamma}}{g_{A_2}}\right)\\
\label{eq:RGratios02}
\frac{dg_{\gamma}}{dg_{A_2}}&=&\mathcal{R}_{42}\left(\frac{g_{A_1}}{g_{A_2}},\frac{g_{D_2}}{g_{A_2}},\frac{g_{\gamma}}{g_{A_2}}\right),
\end{eqnarray}
where
\begin{eqnarray}
\mathcal{R}_{12}\left(x,y,z\right)&=&\frac{4x}{x^2-2x+8
-2y+y^2+4\left(3-y\right)z+6z^2}\nonumber\\\\
\mathcal{R}_{32}\left(x,y,z\right)&=&-\frac{4\left(1+xy-3y-y^2+4z+z^2\right)}{x^2-2x+8
-2y+y^2+4\left(3-y\right)z+6z^2}\nonumber\\\\
\mathcal{R}_{42}\left(x,y,z\right)&=&\frac{2z\left(-2x+2+2z\right)}{x^2-2x+8
-2y+y^2+4\left(3-y\right)z+6z^2}.\nonumber\\
\end{eqnarray}
Equations (\ref{eq:RGratios00}-\ref{eq:RGratios02}) are homogeneous,
which means that we can instead study the flow of the coupling
constant ratios
\begin{eqnarray}
\label{eq:RGratios10}
g_{A_2}\frac{d\frac{g_{A_1}}{g_{A_2}}}{dg_{A_2}}&=&-\frac{g_{A_1}}{g_{A_2}}+\mathcal{R}_{12}\left(\frac{g_{A_1}}{g_{A_2}},\frac{g_{D_2}}{g_{A_2}},\frac{g_{\gamma}}{g_{A_2}}\right)\\
\label{eq:RGratios11}
g_{A_2}\frac{d\frac{g_{D_2}}{g_{A_2}}}{dg_{A_2}}&=&-\frac{g_{D_2}}{g_{A_2}}+\mathcal{R}_{32}\left(\frac{g_{A_1}}{g_{A_2}},\frac{g_{D_2}}{g_{A_2}},\frac{g_{\gamma}}{g_{A_2}}\right)\\
\label{eq:RGratios12}
g_{A_2}\frac{d\frac{g_{\gamma}}{g_{A_2}}}{dg_{A_2}}&=&-\frac{g_{\gamma}}{g_{A_2}}+\mathcal{R}_{42}\left(\frac{g_{A_1}}{g_{A_2}},\frac{g_{D_2}}{g_{A_2}},\frac{g_{\gamma}}{g_{A_2}}\right).
\end{eqnarray}
Note that the right hand side of these equations is a function of
coupling constant {\it ratios only}, i.e. it is autonomous in the
new variables
$\frac{g_{A_1}}{g_{A_2}},\frac{g_{D_2}}{g_{A_2}},\frac{g_{\gamma}}{g_{A_2}}$.
We can think of the right hand sides effectively as (highly
non-linear) $\beta$-functions for the ratios. The advantage of
rewriting the flow equations this way, is that in this form it is
easier to analyze the qualitative nature of the flow diagram. Unlike
in the case where $g_{\gamma}$ was assumed to vanish from the
start\cite{VafekYang2010PhysRevB.81.041401}, in the present case the
$\beta$-function for $g_{A_2}$ is not negative semidefinite. It may
appear therefore, that we lose the directionality of the flow
equations in the three dimensional ratio space. This turns out not
to be the case, since the (ellipsoidal) region in the 3D ratio space
where $dg_{A_2}/d\ln s$ changes sign is precisely the same region
where the "$\beta$"-functions for the ratios change sign, and so it
is enough to determine the directionality of the flow of the
trajectories near fixed points of the ratios, which turns out to be
simple enough.

The qualitative analysis proceeds by finding the fixed points in the
ratio space. There are four of them:
\begin{eqnarray}\label{eq:fixedRatios1}
\left(\frac{g^*_{A_1}}{g^*_{A_2}},\frac{g^*_{D_2}}{g^*_{A_2}},\frac{g^*_\gamma}{g^*_{A_2}}\right)&=&\left(0,-1.085,0\right)\\
\label{eq:fixedRatios2}
\left(\frac{g^*_{A_1}}{g^*_{A_2}},\frac{g^*_{D_2}}{g^*_{A_2}},\frac{g^*_\gamma}{g^*_{A_2}}\right)&=&\left(0,0.566,0\right)\\
\label{eq:fixedRatios3}
\left(\frac{g^*_{A_1}}{g^*_{A_2}},\frac{g^*_{D_2}}{g^*_{A_2}},\frac{g^*_\gamma}{g^*_{A_2}}\right)&=&\left(0,6.519,0\right)\\
\label{eq:fixedRatios4}
\left(\frac{g^*_{A_1}}{g^*_{A_2}},\frac{g^*_{D_2}}{g^*_{A_2}},\frac{g^*_\gamma}{g^*_{A_2}}\right)&=&\left(-1,-1,-1\right)
\end{eqnarray}
The first three are the $N=2$ analog of the ($N=4$) fixed ratios
found in Ref.\cite{VafekYang2010PhysRevB.81.041401}, while the
fourth one is new. For $g_{A_2}<0$ the stability analysis gives the
first (\ref{eq:fixedRatios1}) and the third one
(\ref{eq:fixedRatios3}) to be sinks (see
Fig.\ref{fig:BilayerRGflow}). The second one (\ref{eq:fixedRatios2})
is mixed, with two stable directions (negative eigenvalues) and one
unstable direction (positive eigenvalue). The fourth one
(\ref{eq:fixedRatios4}) is also mixed, with one positive, one
negative and one zero eigenvalue. For $g_{A_2}>0$ the directionality
of the flows is reversed and the sinks become sources while the
mixed fixed points remain mixed but also with reversed sense of
flow. These "runaway" flows in the coupling constant ratio space for
$g_{A_2}>0$ simply correspond to decrease of $g_{A_2}$ which
eventually crosses zero, where the ratios become infinite, and then
become negative. Once negative, the flows are described by the two
stable sinks, separated by a critical plane (3D version of the (red)
separatrix shown in Fig.3 of
Ref.\cite{VafekYang2010PhysRevB.81.041401}).

The generic flow for initial $g_{A_2}$ of any sign is towards large
and negative $g_{A_2}$ and towards either one of the two ratio
sinks.

However, it is interesting to ask, under which (non-trivial)
conditions, may all the coupling constants flow to zero. One
possibility, is to fine-tune the initial values of $g_{A_1}$ and
$g_{\gamma}$ to zero, set the initial value of $g_{A_2}>0$ and the
ratio $-1.085\leq g_{D_2}/g_{A_2}\leq 6.519$. In this case, the flow
is towards both $g_{A_2}\rightarrow 0$ and $g_{D_2}\rightarrow 0$
while their ratio approaches $0.566$. Note that in this case we have
to fine-tune two of the four symmetry allowed couplings
$g_{A_1},g_{\gamma}$ to vanish.

Another possibility involves the new fixed point in the ratio space
at
$\left(\frac{g_{A_1}}{g_{A_2}},\frac{g_{D_2}}{g_{A_2}},\frac{g_\gamma}{g_{A_2}}\right)=(-1,-1,-1)$.
 While the fixed point is mixed, in that
one of the RG eigenvalues is negative and one positive, one
 eigenvalue, whose right eigenvector is $(\frac{1}{\sqrt{3}},\frac{1}{\sqrt{3}},\frac{1}{\sqrt{3}})$, vanishes.
This means that in the vicinity of this fixed point, the flow along
or against this direction in the ratio space is very slow.
Importantly, our numerical integration finds that for $g_{A_2}>0$,
starting {\it anywhere}(!) along the line
$\left(\frac{g_{A_1}}{g_{A_2}},\frac{g_{D_2}}{g_{A_2}},\frac{g_\gamma}{g_{A_2}}\right)=(-\lambda,-\lambda,-\lambda)$
for $0<\lambda\leq1$, the flow is towards $\lambda=1$ with {\it
decreasing} $g_{A_2}\rightarrow 0$. The flow trajectories passing
through this line segment in the ratio space, however, are {\it not}
straight lines. In fact, they connect with the fixed point
$\left(\frac{g_{A_1}}{g_{A_2}},\frac{g_{D_2}}{g_{A_2}},\frac{g_\gamma}{g_{A_2}}\right)=(0,-1.085,0)$.
This means that there is a nontrivial (curved) finite surface in the
ratio space along which the flow is directed towards the {\it
non-interacting} fixed point if $g_{A_2}$ starts out positive. In
this case only one parameter needs to be fine-tuned in order to
start on this surface. This interesting behavior, however, is
non-generic, in that such a surface is unstable, and the generic
flow for initial $g_{A_2}>0$ is towards large and negative $g_{A_2}$
and towards the two ratio sinks.

\subsection{Susceptibilities and ordered states}
The physics associated with the fixed ratios analyzed in the
previous section can be understood by studying the flow of the
susceptibilities toward forming various orders. For translationally
invariant order parameters, the susceptibilities can be calculated
from the above flows by introducing source terms into the action so
that $S\rightarrow S+\Delta\mathcal{S}$
\begin{eqnarray}
\Delta\mathcal{S}&=&-\Delta_{ph}^{\mathcal{O}_i}\int d\tau d^2\br
\psi^{\dagger}(\br,\tau)\mathcal{O}^{(i)}\psi(\br,\tau)\nonumber\\
&-&\Delta_{pp}^{\mathcal{O}_i}\int d\tau
d^2\br\psi_{a}(\br,\tau)\mathcal{O}^{(i)}_{ab}\psi_b(\br,\tau)
\end{eqnarray}
Next, we integrate out the fermionic modes within a small shell
given by $\Lambda/s<k<\Lambda$ and find the correction to the source
term perturbatively in the $g$'s. We then substitute the flow of the
$g$'s into the prefactors of various source terms and ask which
diverges the fastest as $s$ increases.

\subsubsection{Particle-hole channels:}
\begin{table}[t]
\begin{center}
    \begin{tabular}{ | c || c | c | c | c |}
    \hline
    $\psi^{\dagger}X_j\psi$ & $1$ & $2$ & $3$ & $4$\\
    \hline\hline
    $\Delta_A$ & $0,0,0,0$  & $1,-1,-4,-2$ & $0,0,0,0$  & $1,-1,0,0$\\ \hline
    $\Delta_B$ & $1,-1,0,2$ & $0,0,0,0$    & $0,0,0,0$  & $1,-1,0,0$ \\ \hline
    $\Delta_C$ & $2,2,-4,-4$ & $1,-1,0,2$    & $2,2,-4,-8$  & $1,-1,0,0$ \\ \hline
    $\Delta_D$ & $1,-1,-4,-2$  & $1,-6,-4,4$   & $2,2,-4,-8$& $1,-1,0,0$\\ \hline
    \hline
    \end{tabular}
\end{center}
\caption{The susceptibility coefficients
$\mathcal{A}_{X_j},\mathcal{B}_{X_j},\mathcal{C}_{X_j},\mathcal{D}_{X_j}$
in Eq.(\protect\ref{eq:vertexGeneral_ph}) for different
particle-hole order parameters $\psi^{\dagger}\mathcal{O}_i\psi$. }
\label{tab:ApBpCp}
\end{table}

In the particle-hole channel, we therefore find:
\begin{eqnarray}
&&\Delta^{\mathcal{O}_i}(s)\psi_<^{\dagger}\mathcal{O}_i\psi_<(\br,\tau)
=
s^2\Delta^{\mathcal{O}_i}(1)\psi^{\dagger}_<\mathcal{O}_i\psi_{<}\nonumber\nonumber\\
&+&s^2\Delta^{\mathcal{O}_i}(1)\sum_{M}g_{MM} \Pi_{\mathcal{O}M}
\psi_{<}^{\dagger}M\psi_<(\br,\tau)\nonumber\\
&-&s^2\Delta^{\mathcal{O}_i}(1)\sum_{M}g_{MM}\psi_<^{\dagger}\Upsilon_{\mathcal{O}M}\psi_<(\br,\tau)
\end{eqnarray}
where $i$ is summed over the $16$ independent order parameters
(generators of $SU(4)$), and
\begin{eqnarray}
\Pi_{\mathcal{O}M}\!&=&\!\!\int^{\infty}_{-\infty}\!\!\frac{d\omega}{2\pi}\int^{\Lambda}_{\Lambda/s}
\!\!\frac{d^2\bk}{(2\pi)^2}\mbox{Tr}\left[G_{\bk}(i\omega)\mathcal{O}_iG_{\bk}(i\omega)M\right]\\
\Upsilon_{\mathcal{O}M}\!&=&\!\!\int^{\infty}_{-\infty}\!\!\frac{d\omega}{2\pi}\int^{\Lambda}_{\Lambda/s}
\!\!\frac{d^2\bk}{(2\pi)^2}M
G_{\bk}(i\omega)\mathcal{O}_iG_{\bk}(i\omega)M
\end{eqnarray}
Using (\ref{eq:greensFxnTrick}) one can easily convince oneself that
the only non-zero contributions to $\Pi_{\mathcal{O}M}$ come from
$\mathcal{O}=M$, and that the matrix $\Upsilon_{\mathcal{O}M}$ is
proportional to $\mathcal{O}$. From here we find the flow equations
for the source terms
\begin{eqnarray}\label{eq:vertexGeneral_ph}
\frac{d\ln\Delta_{X_j}}{d\ln
s}\!\!\!&=&\!\!2\!+\!\left(\mathcal{A}_{X_j}g_{A_1}+\mathcal{B}_{X_j}g_{D_2}+\mathcal{C}_{X_j}g_{A_2}+\mathcal{D}_{X_j}g_{\gamma}\right)\frac{m}{4\pi}\nonumber\\
\end{eqnarray}
where $X=A,B,C,D$ and $j=1,2,3,4$. The results of this calculation,
i.e. the values of
$\mathcal{A}_{X_j},\mathcal{B}_{X_j},\mathcal{C}_{X_j}$, and
$\mathcal{D}_{X_j}$, are shown in Table \ref{tab:ApBpCp}. The
coupling constants $g$ are functions of $s$ and, in order to
determine the most likely ordering tendency, it is necessary to find
out which source term $\Delta_{X_j}$ grows the fastest. We can write
each of these equations as
\begin{eqnarray}\label{eq:vertexGeneral_ph_1}
\frac{d\ln\Delta_{X_j}}{d\ln
s}\!&=&\!\!2+g_{A_2}\left(\mathcal{A}_{X_j}\frac{g_{A_1}}{g_{A_2}}+\mathcal{B}_{X_j}\frac{g_{D_2}}{g_{A_2}}\right.\nonumber\\
&+&\left.\mathcal{C}_{X_j}+\mathcal{D}_{X_j}\frac{g_{\gamma}}{g_{A_2}}\right)\frac{m}{4\pi}
\end{eqnarray}
and near the two sinks, we can take $g_{A_2}<0$ and substitute the
fixed point ratios.

Near the first sink
$\left(\frac{g^*_{A_1}}{g^*_{A_2}},\frac{g^*_{D_2}}{g^*_{A_2}},\frac{g^*_\gamma}{g^*_{A_2}}\right)=\left(0,-1.085,0\right)$
and, plugging in these values, we find that the fastest divergence
appears for $\Delta^{C_1}_{ph}$, $\Delta^{C_3}_{ph}$, and
$\Delta^{D_3}_{ph}$. Discriminating between the first and the last
two ordering tendencies requires knowledge of the sign of the
subleading ratio $\frac{g_\gamma}{g_{A_2}}$.

If $\frac{g_\gamma}{g_{A_2}}<0$, i.e. if it approaches $0$ from
below as $s$ increases, then the most dominant particle-hole
ordering tendency is towards a finite expectation value of
$C_1=\gamma_0=1_2\sigma_z$. Physically, this order parameter, which
corresponds to an imbalance in the number of particles on the two
different layers, opens up a gap at the $\bK$ and $-\bK$ points in
the Brillouin zone and the system is a (trivial) insulator. As shown
below, this turns out to be the case for a lattice model with a
nearest neighbor repulsion $V$.

On the other hand, if $\frac{g_\gamma}{g_{A_2}}>0$ and approaches
zero from above as $s$ increases, then the most dominant ordering
tendency among the particle-hole channels studied here is towards
finite expectation values of $C_3=\gamma_3=\tau^x\sigma^y$ and
$D_3=\gamma_5=\tau^y\sigma^y$, both of which are odd under time
reversal symmetry (\ref{eq:timeReversal}).

Near the second sink
$\left(\frac{g^*_{A_1}}{g^*_{A_2}},\frac{g^*_{D_2}}{g^*_{A_2}},\frac{g^*_\gamma}{g^*_{A_2}}\right)=\left(0,6.519,0\right)$
the most dominant ordering tendency is towards a finite expectation
value of $D_2=i\gamma_1\gamma_2=\tau^z\sigma^z$. The corresponding
order parameter also opens up a gap in the single particle spectrum,
but unlike $C_1$, it breaks time reversal symmetry. This results in
an {\it anomalous quantum Hall state}, with zero B-field Hall
conductivity $\sigma_{xy}=\pm 2\frac{e^2}{h}$. Such a state is a
bilayer analog of the Haldane model for the quantum Hall effect
without Landau levels\cite{HaldanePRL1988PhysRevLett.61.2015}.

\subsubsection{Particle-particle channels:} Since our Fermions are
spinless, if the integral
$$
\int d^2\br\psi_{a}(\br,\tau)\mathcal{O}^{(i)}_{ab}\psi_b(\br,\tau)
$$
is to be finite, we must have
$\mathcal{O}_{ab}^{i}=-\mathcal{O}_{ba}^{i}$. Of the sixteen SU(4)
generators (\ref{eq:su4generators}) this condition selects the six
matrices $B_3,B_4,C_2,C_3,D_1$, and $D_4$. Integrating out the fast
modes, we are left with the following renormalization of the source
term for the slow modes $\psi_{<}$
\begin{eqnarray}
&&\Delta^{\mathcal{O}_i}(s)\psi^{<}_{a}(\br,\tau)\mathcal{O}^{(i)}_{ab}\psi^{<}_b(\br,\tau)=
s^2\Delta^{\mathcal{O}_i}(1)\psi^{<}_{a}\mathcal{O}^{(i)}_{ab}\psi^{<}_b(\br,\tau)\nonumber\\
&+&s^2\Delta^{\mathcal{O}_i}(1)\sum_{M}g_{MM}\int\frac{d\omega}{2\pi}\int^{\Lambda}_{\Lambda/s}
\frac{d^2\bk}{(2\pi)^2}\nonumber\\
&&\mathcal{O}_{\alpha\beta}^{i}G_{\bk,\beta
a}(i\omega_n)M_{ab}\psi_{b<}(\br,\tau) G_{-\bk,\alpha
c}(i\omega_n)M_{cd}\psi_{d<}(\br,\tau)\nonumber\\
\end{eqnarray}
Evaluating the necessary matrix products leads to
\begin{eqnarray}\label{eq:susceptibiltySpinlessB3}
\frac{d\ln\Delta_{B_3}}{d\ln s}&=&2-\left(g_{A_1}+g_{D_2}-2g_{\gamma}\right)\frac{m}{4\pi}\\
\frac{d\ln\Delta_{B_4}}{d\ln s}&=&2-\left(2g_{A_1}-2g_{D_2}+4g_{A_2}+4g_{\gamma}\right)\frac{m}{4\pi}\\
\frac{d\ln\Delta_{C_3}}{d\ln s}&=&2-\left(g_{A_1}+g_{D_2}-2g_{\gamma}\right)\frac{m}{4\pi}\\
\label{eq:susceptibiltySpinlessC2} \frac{d\ln\Delta_{C_2}}{d\ln
s}&=&\frac{d\ln\Delta_{D_1}}{d\ln s}=\frac{d\ln\Delta_{D_4}}{d\ln
s}=2.
\end{eqnarray}
Near the first sink,
$\left(\frac{g^*_{A_1}}{g^*_{A_2}},\frac{g^*_{D_2}}{g^*_{A_2}},\frac{g^*_\gamma}{g^*_{A_2}}\right)=\left(0,-1.085,0\right)$,
and, substituting these values into
Eqs.(\ref{eq:susceptibiltySpinlessB3})-(\ref{eq:susceptibiltySpinlessC2}),
we find that the strongest divergence appears for
$\Delta^{B_4}_{pp}$. The {\it leading} divergence is as fast as for
$\Delta^{C_1}_{ph}$, $\Delta^{C_3}_{ph}$, and $\Delta^{D_3}_{ph}$,
but it differs in the subleading terms. In fact, for $g_{A_1}>0$,
the strongest divergence is in the particle-hole channel
$\Delta^{C_1}_{ph}$ discussed above. In principle, fine-tuning and
keeping the subleading term $g_{A_1}/g_{A_2}<0$ (and setting
$g_{\gamma}=0$ or keeping $g_{\gamma}/g_{A_2}<0$) may lead to the
strongest divergence appearing in the particle-particle channel.

\section{$t-V$ model for spinless fermions}
While the weak coupling results are quite general, we can apply them
to a specific microscopic model, which happens to be quite revealing
in that we can also analyze it for strong coupling and thus compare
the two regimes. We consider spinless fermions hopping on the
half-filled A-B stacked honeycomb bilayer, with nearest neighbor
hoppings $t$ and $t_{\perp}$ only and with nearest neighbor
repulsions $V$ and $V_{\perp}$. The corresponding Hamiltonian is
\begin{eqnarray}\label{eq:tV0}
\mathcal{H}&=&H^{\perp}_0+H_0^{\parallel}+\mathcal{V}^{\perp}+\mathcal{V}^{\parallel}
\end{eqnarray}
where
\begin{eqnarray}\label{eq:tV1}
H^{\perp}_0&=&-t_{\perp}\sum_{\bR}\left(a^{\dagger}_{1}(\bR)a_{2}(\bR)+h.c.\right)\\
H_0^{\parallel}&=&-t\sum_{\bR,\delta}\left(b^{\dagger}_{1}(\bR+\delta)a_{1}(\bR)+b^{\dagger}_{2}(\bR-\delta)a_{2}(\bR)+h.c.
\right)\nonumber\\
\mathcal{V}^{\perp}&=&V_{\perp}\sum_{\bR}
\left(a^{\dagger}_{1}(\bR)a_{1}(\bR)-\frac{1}{2}\right)\left(a^{\dagger}_{2}(\bR)a_{2}(\bR)-\frac{1}{2}\right)\nonumber\\
\mathcal{V}^{\parallel}&=&V\sum_{\bR,\delta}\left(a^{\dagger}_{1}(\bR)a_{1}(\bR)-\frac{1}{2}\right)\left(b^{\dagger}_{1}(\bR+\delta)b_{1}(\bR+\delta)-\frac{1}{2}\right)\nonumber\\
&+&V\sum_{\bR,\delta}\left(a^{\dagger}_{2}(\bR)a_{2}(\bR)-\frac{1}{2}\right)\left(b^{\dagger}_{2}(\bR-\delta)b_{2}(\bR-\delta)-\frac{1}{2}\right)\nonumber\\
\end{eqnarray}

\subsection{Weak coupling}
In order to project onto the low energy modes, we first rewrite the
Hamiltonian (\ref{eq:tV0}) as an imaginary time Grassman path
integral. We then integrate out the $a_1$ and $a_2$ modes
perturbatively. This results in
\begin{eqnarray}\label{eq:tVLeff}
\mathcal{L}^{(t-V)}_{eff}=\mathcal{L}_{0}+\mathcal{L}_{int}
\end{eqnarray}
 where
\begin{eqnarray}\label{eq:tVL0}
\mathcal{L}_{0}&=&\frac{t^2}{t_{\perp}}\sum_{\bR,\delta,\delta'}\left(b^{\dagger}_1(R+\delta,\tau)b_2(R-\delta',\tau)+h.c.\right)
\\\label{eq:tVLint}
\mathcal{L}_{int}&=&-\frac{V^2}{8t_{\perp}}\sum_{\bR}
\left(\sum_{\delta}n_{b_1}(\bR+\delta,\tau)-n_{b_2}(\bR-\delta,\tau)\right)^2.\nonumber\\
\end{eqnarray}
Fourier transforming the Fermi modes in the first term
(\ref{eq:tVL0}) gives rise to the kinetic energy term
(\ref{eq:Lag0}) with $m=2t_{\perp}/(9t^2)$. The interaction term can
be written as
\begin{eqnarray}
\mathcal{L}_{int}&=&-\frac{V^2}{8t_{\perp}}\frac{1}{N_{uc}}\sum_{\bq,\bG}
\left[\left(d_{-\bq}n_{b_1,-\bq}(\tau)-d_{\bq}n_{b_2,-\bq}(\tau)\right)\times\right.\nonumber\\
&&\left.\left(d_{\bq+\bG}n_{b_1,\bq+\bG}(\tau)-d_{\bG-\bq}n_{b_2,\bG+\bq}(\tau)\right)\right]
\end{eqnarray}
where $\bG$ is a reciprocal lattice vector and
$n_{b_{j,\bq}}=\frac{1}{N_{uc}}\sum_{\bk}b^{\dagger}_{j,\bk}b_{j,\bk+\bq}$.
In addition, each fermionic mode is restricted to reside in the
first Brillouin zone. Taking $\bk$ in the above sum to be near $\bK$
or $-\bK$ gives two possibilities for $\bq$: either $\bq\sim 0$ or
$\bq\sim \pm2\bK$. Note that in the first case $d_{0}=3$ while in
the second case $d_{\pm2\bK}=0$. Therefore, only the first term
contributes a marginal coupling, and the above Hamiltonian gives
rise to the low energy interaction Lagrangian
\begin{eqnarray}
\mathcal{L}^{t-V}_{int}=g^{(0)}_{C_1}\int d^2\br
(\psi^{\dagger}C_1\psi(\br,\tau))^2
\end{eqnarray}
where
\begin{eqnarray}
g^{(0)}_{C_1}=-\frac{9V^2}{8t_{\perp}}A^{-1}_{uc}.
\end{eqnarray}
The area of the unit cell
$A_{uc}=\hat{z}\cdot(\bR_1\times\bR_2)=\frac{3\sqrt{3}}{2}a^2$. This
means that we should start our RG flow with a small and negative
(attractive) $g^{(0)}_{C_1}(\psi^{\dagger}C_1\psi)^2$, which should
be rewritten using the Fierz identity (\ref{eq:FierzC1}). The
initial conditions are therefore
\begin{eqnarray}
g_{A_1}(s=1)&=&-g^{(0)}_{C_1},\;\;\;g_{D_2}(s=1)=-2g^{(0)}_{C_1}\\
g_{A_2}(s=1)&=&g^{(0)}_{C_1},\;\;\;g_{\gamma}(s=1)=-g^{(0)}_{C_1}.
\end{eqnarray}
Substituting these as the initial conditions into our RG equations
we find that none of the coupling constants change sign and they all
diverge at the same value of $s$. The ratios of the couplings flow
to the fixed point
$\left(\frac{g^*_{A_1}}{g^*_{A_2}},\frac{g^*_{D_2}}{g^*_{A_2}},\frac{g^*_\gamma}{g^*_{A_2}}\right)=\left(0,-1.085,0\right)$.
Therefore, as discussed in the previous section, the fastest
divergence appears in the channel $1_2\sigma_z$. We therefore
conclude that the weak coupling instability of this model is towards
a gapped, broken inversion symmetry state with an imbalance of the
number of particles on layer 1 compared to layer 2.

\subsection{Strong coupling limit}
\begin{figure}[t]
\begin{center}
\includegraphics[width=0.5\textwidth]{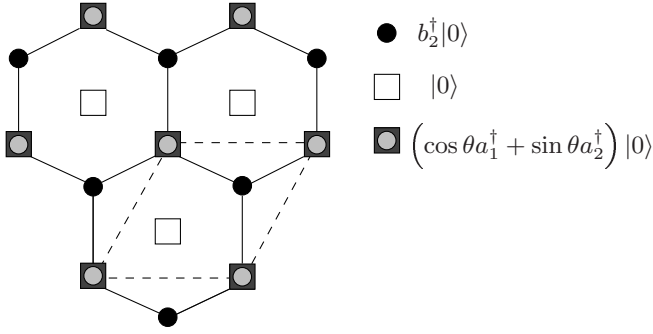}
\end{center}
  \caption{Schematic representation of the strong coupling state for
the spinless $t-V$ model. For $t=0$ but finite $t_{\perp}$, $V$, and
$V_{\perp}$, the sublattice $b_1$ is empty, while sublattice $b_2$
is fully occupied. The $a_1-a_2$ dimer is singly occupied, with an
electron partially {\it delocalized} onto $a_2$ despite the
repulsion from the occupied $b_2$ sites. Such gapped state breaks
inversion symmetry between the layers $1$ and $2$.
  }\label{fig:StrongCoupling}
\end{figure}

Setting $t=t_{\perp}=0$ we find three ground states at half-filling:\\
i) each site of sublattice $a_1$ and of $b_2$ is singly occupied\\
ii) each site of sublattice $a_2$ and of $b_1$ is singly occupied\\
iii) each site of sublattice $b_1$ and of $b_2$ is singly occupied.

Each of these states breaks sublattice symmetry, but the average
density of particles on each layer is the same and equal to $1$ per
unit cell (which contains two sites in each layer). The states i)
and ii) differ from the state iii) by the occupation of the
$a_1$-$a_2$ dimer which is singly occupied for the former and empty
for the latter.

If we now set $t=0$ but $t_{\perp}\neq 0$ then we can further lower
the energy of i) and ii) by delocalizing the electron on the dimer.
So, consider the deformation of the state i): we seek a state of the
form
\begin{eqnarray}
|\Psi_{\theta}\rangle&=&\prod_{\bR} (\cos\theta
a_1^{\dagger}(\bR)+\sin\theta
a_2^{\dagger}(\bR))b_2^{\dagger}(\bR+\delta)|0\rangle
\end{eqnarray}
For $t_{\perp}=0$, we have $\theta=0$, but once $t_{\perp}\neq 0$ we
expect $\theta\neq 0$.

Acting on $|\Psi_{\theta}\rangle$ with $\mathcal{H}$ (for $t=0$) and
requiring $|\Psi_{\theta}\rangle$ to be an eigenstate gives:
\begin{eqnarray}
\left(-\frac{V_{\perp}}{4}-\frac{3V}{2}\right)\cos\theta-t_{\perp}\sin\theta
&=& E\cos\theta\\
-t_{\perp}\cos\theta+\left(-\frac{V_{\perp}}{4}+\frac{3V}{2}\right)\cos\theta
&=& E\sin\theta
\end{eqnarray}
The above equation has two eigenvalues
$E_{\pm}=-\frac{V_{\perp}}{4}\pm\sqrt{t_{\perp}^2+\frac{9V^2}{4}}$,
and clearly $t_{\perp}$ favors a state with a delocalized particle
on the dimer. Thus in the ground state
\begin{eqnarray}
\cos\theta&=&\frac{1}{\sqrt{2}}\sqrt{1+\frac{3V}{2\left(E_++\frac{V_{\perp}}{4}\right)}}\\
\sin\theta&=&\frac{1}{\sqrt{2}}\sqrt{1-\frac{3V}{2\left(E_++\frac{V_{\perp}}{4}\right)}}.
\end{eqnarray}
This state breaks the sublattice symmetry and there are clearly more
particles on layer $2$ than on layer $1$. Similarly, if we deform
ii) in analogous way, we will find a state with more particles on
layer $1$ than on layer $2$. Both of these states are gapped.

For infinitesimal $t$, we expect the energy of the broken symmetry
state to be further lowered via second-order processes. This leads
us to the conclusion that in the strong coupling limit, our
Hamiltonian $\mathcal{H}$ has a ground state with broken inversion
symmetry, i.e. the total number of particles on the upper layer is
different from the total number of particles on the lower layer.

\section{Spin-$\frac{1}{2}$ case}
The symmetry based reduction of the number of coupling can be used
for the spin-$\frac{1}{2}$ case as well. All the arguments presented
in the section dealing with short range interactions follow through,
but now the Fierz vector is $18$-dimensional, instead of $9$.
Specifically, each term in Eq. (\ref{eq:FierzVector}), when
multiplied by the appropriate coupling, gives rise to two terms as
\begin{eqnarray}
g_{X_j}(\psi^{\dagger}X_j\psi)^2\rightarrow
g^{(c)}_{X_j}(\psi_{\alpha}^{\dagger}X_j\psi_{\alpha})^2+g^{(s)}_{X_j}(\psi_{\alpha}^{\dagger}X_j\vec{\sigma}_{\alpha\beta}\psi_{\beta})^2,\nonumber\\
\end{eqnarray}
where the Pauli $\vec{\sigma}$ corresponds to spin-$\frac{1}{2}$
$SU(2)$. The seemingly independent couplings in the two different
channels, $c$ and $s$, are still related to each other via a
Fierz-like identity.

In particular, we can use the $SU(8)$ algebraic identity
\begin{eqnarray}
S_{ij}T_{mn}
&=&\frac{1}{64}\mbox{Tr}[S\Lambda^aT\Lambda^b]\Lambda^{b}_{in}\Lambda^a_{mj}
\end{eqnarray}
where $S$ and $T$ are $8\times8$ matrices, and the $64$ generators
$\Lambda^a$ can be obtained from the $16$-$SU(4)$ generators as
$\{\Gamma^a\otimes 1,\Gamma^a\otimes \sigma^z, \Gamma^a\otimes
\sigma^x,\Gamma^a\otimes \sigma^y\}$. This leads to
\begin{eqnarray}\label{eq:Fierz8by8}
&&\left(\psi^{\dagger}(x)S\psi(x)\right)\left(\psi^{\dagger}(y)T\psi(y)\right)=\nonumber\\
&&-\frac{1}{64}\mbox{Tr}[S\Lambda^aT\Lambda^b]\left(\psi^{\dagger}(x)\Lambda^b\psi(y)\right)\left(\psi^{\dagger}(y)\Lambda^a\psi(x)\right).
\end{eqnarray}
Again, the minus sign comes from $\psi$ and $\psi^{\dagger}$ being
anti-commuting (four component) Grassman fields. For contact terms
$x=y$ and the above equation constitutes a set of linear relations
among the $18$ symmetry allowed terms.

If we now arrange the quartic terms into an $18$-component vector
$\mathcal{V}$ (Eq.\ref{eq:FierzVectorSpinHalf}) we can write the
above constraint as
\begin{eqnarray}
\mathcal{F}\mathcal{V}&=&0
\end{eqnarray}
where the matrix $\mathcal{F}$, displayed in Appendix B
(\ref{eq:FierzMatrixSpinHalf}), has nine zero eigenvalues, and, as a
result\cite{HerbutJuricicRoy2009PhysRevB.79.085116} there are $9$
independent couplings in the spin-$\frac{1}{2}$ case. From here it
can be shown that one can eliminate {\it all} the (spin-spin)
$g_X^{s}$ couplings in favor of the (charge-charge) $g_X^{c}$
couplings\cite{Lemonik2010arXiv1006.1399L}.

Using the same technique as described for the spinless case in the
Appendix, we find the RG flow equations for the nine couplings in
the spin-$1/2$ case. These equations,
(\ref{eq:RGspinfulA1}-\ref{eq:RGspinfulgamma}), are shown explicitly
at the end of the Appendix A. While full analysis of the
Eqs.(\ref{eq:RGspinfulA1}-\ref{eq:RGspinfulgamma}) is beyond the
scope of this paper, we have studied the effect of three of these
couplings in Ref.\cite{VafekYang2010PhysRevB.81.041401}, starting
with $g^{(c)}_{A_1}$, and generating $g^{(c)}_{A_2}$ and
$g^{(c)}_{D_2}$. No other couplings are generated, assuming they
vanish to begin with, in agreement with Eqs.
(\ref{eq:RGspinfulA1}-\ref{eq:RGspinfulgamma}). The equations
presented in the Appendix of this paper reduce to Eqs.(6-8) of
Ref.\cite{VafekYang2010PhysRevB.81.041401} provided we set $N=4$
there and identify $g_1\leftrightarrow g^{(c)}_{A_1}$,
$g_2\leftrightarrow g^{(c)}_{D_2}$ and $g_3\leftrightarrow
g^{(c)}_{A_2}$. In this case, for finite initial $g^{(c)}_{A_1}>0$
and vanishing initial $g^{(c)}_{A_2}$ and $g^{(c)}_{D_2}$, the most
dominant divergence appears in the nematic channel, which
corresponds to one of the ratio sinks
$g*^{(c)}_{D_2}/g*^{(c)}_{A_2}=m_1\approx -0.525$. For certain
combinations of the couplings, a different sink
($g*^{(c)}_{D_2}/g*^{(c)}_{A_2}=m_3\approx 13.98$, top fixed point
in Fig.3 in Ref.\cite{VafekYang2010PhysRevB.81.041401}) may be
reached. Thus, an anomalous quantum Hall state may in principle be
stabilized in weak coupling as well.
\begin{widetext}
\begin{table}[t]
\begin{center}
    \begin{tabular}{ | c || c | c | c | c |}
    \hline
    $\psi^{\dagger}X_j\otimes 1\psi$ & $1$ & $2$ & $3$ & $4$\\
    \hline\hline
    $\Delta_A$ & $0, 0, 0, 0, 0, 0, 0, 0, 0$  & $1, -1, -8, -2, 0, 1, -1, 2, 0$ & $0, 0, 0, 0, 0, 0, 0, 0, 0$  & $1, -1, 0, 0, 0, -1, 1, 0, -8$\\ \hline
    $\Delta_B$ & $1, -1, 0, 2, -8, 1, -1, -2, 0$ & $0, 0, 0, 0, 0, 0, 0, 0, 0$    & $0, 0, 0, 0, 0, 0, 0, 0, 0$  & $1, -1, 0, 0, 0, -1, 1, 0, -8$ \\ \hline
    $\Delta_C$ & $2, 2, -4, -4, -4, 2, -14, -4, 8$ & $1, -1, 0, 2, -8, 1, -1, -2, 0$    & $2, 2, -4, -16, 4, -2, -2, 0, 0$  & $1, -1, 0, 0, 0, -1, 1, 0, -8$ \\ \hline
    $\Delta_D$ & $1, -1, -8, -2, 0, 1, -1, 2, 0$  & $2, -14, -4, 4, -4, 2, 2, 4, -8$   & $2, 2, -4, -16, 4, -2, -2, 0, 0$& $1, -1, 0, 0, 0, -1, 1, 0, -8$\\ \hline
    \hline
    $\psi^{\dagger}X_j\otimes \vec{\sigma}\psi$ & $1$ & $2$ & $3$ & $4$\\
    \hline\hline
    $\Delta_A$ & $0, 0, 0, 0, 0, 0, 0, 0, 0$  & $1, -1, 0, -2, 0, 1, -1, 2, 0$ & $0, 0, 0, 0, 0, 0, 0, 0, 0$  & $1, -1, 0, 0, 0, -1, 1, 0, 0$\\ \hline
    $\Delta_B$ & $1, -1, 0, 2, 0, 1, -1, -2, 0$ & $0, 0, 0, 0, 0, 0, 0, 0, 0$    & $0, 0, 0, 0, 0, 0, 0, 0, 0$  & $1, -1, 0, 0, 0, -1, 1, 0, 0$ \\ \hline
    $\Delta_C$ & $2, 2, -4, -4, -4, 2, 2, -4, 8$ & $1, -1, 0, 2, 0, 1, -1, -2, 0$    & $2, 2, -4, 0, 4, -2, -2, 0, 0$  & $1, -1, 0, 0, 0, -1, 1, 0, 0$ \\ \hline
    $\Delta_D$ & $1, -1, 0, -2, 0, 1, -1, 2, 0$  & $2, 2, -4, 4, -4, 2, 2, 4, -8$   & $2, 2, -4, 0, 4, -2, -2, 0, 0$& $1, -1, 0, 0, 0, -1, 1, 0, 0$\\ \hline
    \hline
    \end{tabular}
\end{center}
\caption{The susceptibility coefficients (spin-$1/2$ case)
$\mathcal{A}_{X_j},\mathcal{B}_{X_j},\mathcal{C}_{X_j},\mathcal{D}_{X_j},\mathcal{E}_{X_j},\mathcal{F}_{X_j},\mathcal{G}_{X_j},\mathcal{H}_{X_j}$,
and $\mathcal{I}_{X_j}$ in Eq.(\protect\ref{eq:vertexGeneral_ph})
for different particle-hole order parameters
$\psi^{\dagger}\mathcal{O}_i\psi$. } \label{tab:spinHalf}
\end{table}
\end{widetext}

Finally, we note in passing that the number (9) of independent
couplings in the spin-$\frac{1}{2}$ case is in agreement with
Ref.\cite{Lemonik2010arXiv1006.1399L}, but disagreement with
Ref.\cite{FanZhangPRB2010PhysRevB.81.041402}.

\subsection{Susceptibilities}
Just as in the spinless case we can analyze the flow of various
source terms $\Delta$ in order to determine the most dominant weak
coupling ordering tendencies. Since in the spin-$1/2$ case there are
9 independent coupling constants, we have
\begin{eqnarray}\label{eq:susceptibiltySpinHalf}
&&\frac{d\ln\Delta_{X_j}}{d\ln
s}\!=\!2\!+\!\left(\mathcal{A}_{X_j}g_{A_1}+\mathcal{B}_{X_j}g_{D_2}+\mathcal{C}_{X_j}g_{A_2}+\mathcal{D}_{X_j}g_{\gamma}
\right.\nonumber\\
&&\left.+\mathcal{E}_{X_j}g_{B_1}+\mathcal{F}_{X_j}g_{B_2}+\mathcal{G}_{X_j}g_{C_1}+\mathcal{H}_{X_j}g_{\alpha}+\mathcal{I}_{X_j}g_{\beta}
\right)\frac{m}{4\pi}.\nonumber\\
\end{eqnarray}
The coefficients $\mathcal{A}-\mathcal{I}$ in 32 different
particle-hole channels are listed in Table \ref{tab:spinHalf}. The
most dominant instability channel, $X_j$, yields the largest
right-hand side of the above equation.

\section{Hubbard model on the A-B stacked honeycomb bilayer}
In this section we use the above machinery to study the weak and
strong coupling limits of the repulsive Hubbard model on the A-B
stacked honeycomb bilayer. Just as before, we assume nearest
neighbor hopping only, and the potential energy term can be written
as
\begin{eqnarray}
\mathcal{V}^{(H)}&=&U\sum^2_{j=1}\sum_{\bR}a^{\dagger}_{j\uparrow}(\bR)a_{j\uparrow}(\bR)a^{\dagger}_{j\downarrow}(\bR)a_{j\downarrow}(\bR)\nonumber\\
&+&U\sum_{\bR}b^{\dagger}_{1\uparrow}(\bR+\delta)b_{1\uparrow}(\bR+\delta)b^{\dagger}_{1\downarrow}(\bR+\delta)b_{1\downarrow}(\bR+\delta)\nonumber\\
&+&U\sum_{\bR}b^{\dagger}_{1\uparrow}(\bR-\delta)b_{1\uparrow}(\bR-\delta)b^{\dagger}_{1\downarrow}(\bR-\delta)b_{1\downarrow}(\bR-\delta)\nonumber\\
\end{eqnarray}

\subsection{Weak coupling limit}
\begin{figure}[t]
\begin{center}
\includegraphics[width=0.5\textwidth]{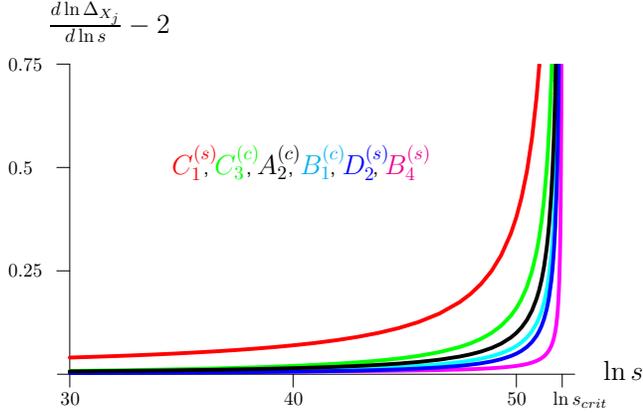}
\end{center}
  \caption{(Color online) Susceptibility vs $\ln s$
  (Eq.\protect\ref{eq:susceptibiltySpinHalf}) for the Hubbard model with initial
  $\frac{m}{4\pi A_c}U=0.01$, in the channels $C^{(s)}_1-B^{(s)}_4$ (left to right). The fastest
divergence appears in the antiferromagnetic channel
$C_1\otimes\vec{\sigma}$. Altogether $32$ particle-hole channels
have been analyzed (Table \protect\ref{tab:spinHalf}); the channels
not shown are either symmetry-related to the ones shown, or $d\ln
\Delta/d\ln s-2$ vanishes (or is
negative).}\label{fig:StrongCoupling}
\end{figure}

Projecting the Hubbard interaction onto the low energy modes we find
\begin{eqnarray}
\mathcal{L}_{int}&=&\frac{U}{2A_{uc}}\sum^2_{j=1}\int  d^2\br
\left[\left(\psi^{\dagger}\mathcal{M}^{(f)}_j\psi\right)
\left(\psi^{\dagger}\mathcal{M}^{(f)}_j\psi\right)\right.\nonumber\\
&+&\left. \left(\psi^{\dagger}\mathcal{M}^{(b)}_j\psi\right)
\left(\psi^{\dagger}{\mathcal{M}^{(b)}_j}^{T}\psi\right)\right.\nonumber\\
&+&\left. \left(\psi^{\dagger}{\mathcal{M}^{(b)}_j}^T\psi\right)
\left(\psi^{\dagger}\mathcal{M}^{(b)}_j\psi\right)\right]
\end{eqnarray}
where
\begin{eqnarray}
\mathcal{M}^{(f)}_1\!\!&=&\!\!\left(\begin{array}{cccc}
1 & 0 & 0 &0 \\
0 & 0 & 0 &0 \\
0 & 0 & 1 &0 \\
0 & 0 & 0 &0
\end{array}\right)\otimes 1_2,\;
\mathcal{M}^{(f)}_2\!\!=\!\!\left(\begin{array}{cccc}
0 & 0 & 0 &0 \\
0 & 1 & 0 &0 \\
0 & 0 & 0 &0 \\
0 & 0 & 0 &1
\end{array}\right)\otimes 1_2,\nonumber\\
\mathcal{M}^{(b)}_{1}\!\!&=&\!\! \left(\begin{array}{cccc}
0 & 0 & 1 &0 \\
0 & 0 & 0 &0 \\
0 & 0 & 0 &0 \\
0 & 0 & 0 &0
\end{array}\right)\otimes 1_2,\;
\mathcal{M}^{(b)}_{2}\!\!=\!\! \left(\begin{array}{cccc}
0 & 0 & 0 &0 \\
0 & 0 & 0 &1 \\
0 & 0 & 0 &0 \\
0 & 0 & 0 &0
\end{array}\right)\otimes 1_2.\nonumber
\end{eqnarray}
The 8-component Fermi fields $\psi$ are understood to be at
space-(imaginary) time point $\br$, $\tau$.

Using the notation established in the previous sections, we note
that the above interaction terms in the low energy effective
Lagrangian can be written as
\begin{eqnarray}\mathcal{L}_{int}&=&\frac{U}{4A_{uc}}\int
d^2\br \left[\left(\psi^{\dagger}A_1\otimes 1\psi\right)^2+\left(\psi^{\dagger}C_1\otimes 1\psi\right)^2\right.\nonumber\\
&+&\left. \frac{1}{2}\sum_{X=A,B,C,D}\left(\psi^{\dagger}X_4\otimes
1\psi\right)^2 \right]
\end{eqnarray}
This means that, of the 9 symmetry-allowed coupling constants, the
only non-zero ones are $g_{A_1}^{(c)}$, $g_{C_1}^{(c)}$ and
$g_{\beta}^{(c)}$, with initial values
\begin{eqnarray}
g_{A_1}^{(c)}(s=1)&=&g_{C_1}^{(c)}(s=1)=\frac{U}{4A_{uc}},\\
g_{\beta}^{(c)}(s=1)&=&\frac{U}{8A_{uc}}.
\end{eqnarray}
Next, we numerically solve the RG flow equations
(\ref{eq:RGspinfulA1})-(\ref{eq:RGspinfulgamma}) with the above
initial conditions (for $mU/(4\pi A_{uc})=0.01$), and substitute the
resulting $s$-dependent couplings into the susceptibility flow
equation (\ref{eq:susceptibiltySpinHalf}) using the coefficients
displayed in Table \ref{tab:spinHalf}. Comparison of the resulting
susceptibilities in 32 particle-hole channels shows that the most
dominant divergence appears for
$\mathcal{O}=C_1\otimes\vec{\sigma}$. Physically, this corresponds
to an anti-ferromagnetic state, with anti-aligned spins on the sites
$b_1$ and $b_2$.

\subsection{Strong coupling}
It is well known that\cite{FradkinBook} in the strong coupling limit
$U/t\gg 1$ the Hubbard model with one particle per site is
equivalent to the spin-$1/2$ Heisenberg model with antiferromagnetic
coupling $J\sim t^2/U$. If we set $t_{\perp}=0$, the two honeycomb
layers decouple and at strong coupling each layer orders
anti-ferromagnetically\cite{Paiva2005PhysRevB.72.085123} with a
sublattice magnetization that is free to point along any direction
on each layer. Once $t_{\perp}$ is finite, the sublattice
magnetizations on the two different layers lock into relative
anti-ferromagnetic arrangement.

We thus find that the half-filled Hubbard model on the A-B stacked
honeycomb bilayer orders antiferromagnetically in both the weak and
strong coupling limits.

\section{Conclusions}
We have studied the effect of short range interactions on fermions
moving on the A-B stacked bilayer. In order to access the "strong
coupling phases"\cite{RaghuKivelsonScalapino2010PhysRevB.81.224505}
from weak coupling, we have fine-tuned the spectrum of the
non-interacting Hamiltonian to achieve parabolic degeneracy, with
the ensuing logarithmically-divergent susceptibilities that appear
in several channels. We have found that, in the spinless fermion
case, the typical dominant ordering tendency opens a spectral gap,
although the nature of the resulting insulating state may be
dramatically different. For example, the weak coupling limit of the
spinless $t-V$ model, with nearest neighbor hopping and nearest
neighbor repulsion, leads to an inversion symmetry breaking
(trivial) insulating phase, while the (right) sink in the RG flow
diagram shown in Fig.(\ref{fig:BilayerRGflow}) corresponds to a
spontaneously time-reversal symmetry breaking anomalous quantum Hall
phase with $\sigma_{xy}=\pm 2e^2/h$. Under certain conditions, the
dominant instability may appear in the particle-particle channel as
well. In addition to the generic instabilities of the spinless
model, fine-tuning of the initial couplings may lead to a flow
towards the non-interacting fixed point. While such behavior is
non-generic, it is interesting that there is an entire surface in
the ratio space (Fig.\ref{fig:BilayerRGflow}) which gives rise to
such a flow for positive initial $g_{A_2}$. We have also studied the
strong coupling limit of the $t-V$ model. In this case, the ground
state wavefunction can be shown to explicitly display inversion
symmetry breaking. Since the same type of order is found in the
asymptotic limits of strong and weak coupling, it is reasonable to
assume that for this specific model, such ordering happens for any
(repulsive) coupling strength.

In the spin-$1/2$ case, we find $9$ independent, symmetry-allowed
couplings, and their RG flow equations. While these equations have
not been studied in their entirety, they reduce to the ones
presented before in Ref.\cite{VafekYang2010PhysRevB.81.041401}, in
which case an analysis similar to the one presented for the spinless
case here, leads to either a nematic phase or an anomalous quantum
Hall phase with $\sigma_{xy}=\pm 4e^2/h$, where the extra factor of
$2$ compared to the spinless case is due to trivial spin degeneracy.

Moreover, these equations
(\ref{eq:RGspinfulA1})-(\ref{eq:RGspinfulgamma}) are solved
numerically for the spin-$1/2$ Hubbard model at half filling. The
initial values of the effective couplings are such that the most
dominant particle-hole instability appears in the anti-ferromagnetic
channel. This dominance has been established by comparing
susceptibilities toward 32 different ordering tendencies. Since the
same instability appears in the strong coupling limit, it is
reasonable to conclude that the antiferromagnetic order sets in for
any $U>0$.

When trigonal warping is taken into account, the logarithmic
infra-red divergences are cut
off\cite{VafekYang2010PhysRevB.81.041401,Lemonik2010arXiv1006.1399L}
by the energy scale corresponding to the deviation from the
parabolic spectrum. This means that the non-interacting system is
stable towards {\it infinitesimal} coupling, i.e. there are no true
weak coupling instabilities. Instead, the interaction strength must
be increased beyond a critical value, which may be quite difficult
to obtain accurately. However, it is worth noting that no such
fine-tuning is necessary for the models with parabolic touching
studied in Ref.\cite{KaiSun2009PhysRevLett.103.046811}, in which
case there are true weak coupling instabilities. Unfortunately, one
cannot just immediately translate the results regarding the
dominance and the nature of the weak coupling broken symmetry states
found here for the honeycomb bilayer, because 1) the location of the
degenerate points in the Brillouin zone is (qualitatively) different
and 2) the lattice symmetry will, in general, allow different
contact terms than those found here.

\begin{acknowledgments} This work was supported in part
by NSF CAREER award grant No. DMR-0955561.
\end{acknowledgments}

\appendix
\section{Details of the RG derivation} For general coupling constants $g_{ST}$, expanding in powers of $g$,
gives the cumulant expansion
\begin{eqnarray}
&&\langle e^{-\frac{1}{2}g_{ST}\int_1
\psi^{\dagger}S\psi\psi^{\dagger}T\psi(1)} \rangle \approx
e^{-\frac{1}{2}g_{ST}\left\langle \int_1
\psi^{\dagger}S\psi\psi^{\dagger}T\psi(1)\right\rangle}\nonumber\\
&\times&\!\!\!
\exp\left[\frac{g_{ST}g_{UV}}{8}\!\!\int_{1,2}\!\!\!\!\left(\left\langle
\left(\psi^{\dagger}S\psi\psi^{\dagger}T\psi(1)\right)
\left(\psi^{\dagger}U\psi\psi^{\dagger}V\psi(2)\right)
\right\rangle \right.\right.\nonumber\\
&-&\left.\left.\left\langle
\left(\psi^{\dagger}S\psi\psi^{\dagger}T\psi(1)\right)\right\rangle
\left\langle\left(\psi^{\dagger}U\psi\psi^{\dagger}V\psi(2)\right)
\right\rangle\right)\right]\nonumber
\end{eqnarray}
where the average $\langle\ldots\rangle$ is with respect to the
gaussian weighting factor. We have used a short-hand $1,2$ for the
modes at the space-(imaginary)time $\tau_{1,2},\br_{1,2}$ and each
$\psi=\psi_{>}+\psi_{<}$. We integrate over the fast modes
$\psi_{>}$ whose wavenumbers $\Lambda/s<k<\Lambda$. The
non-interacting Green's function is
\begin{eqnarray}
G_{\bk}(i\omega)&=&\left(-i\omega+\Sigma\cdot
d_{\bk}\right)^{-1}=\frac{i\omega+\Sigma\cdot
d_{\bk}}{\omega^2+\left(\frac{\bk^2}{2m}\right)^2}
\end{eqnarray}
and just as before $d^x_{\bk}=\frac{k^2_x-k^2_y}{2m}$,
$d^y_{\bk}=-\frac{2k_xk_y}{2m}$, and, in the spinless case,
$\Sigma^x=\gamma_2$ and $\Sigma^y=\gamma_1$.

Using the identities (\ref{eq:greensFxnTrick}), we can evaluate the
needed diagrams. All possible contractions correspond to the
diagrams in the Figure (\ref{fig:graphs}).
\begin{widetext}
For the first diagram we find the following terms
\begin{eqnarray}\label{eq:RGrpa}
\Delta S^{(RPA)}_{eff}&=&\frac{1}{2}\sum_{S\in
\mathcal{G}}\sum_{U\in
\mathcal{G}}g_{S}g_{U}\int_1\int_2\psi^{\dagger}(1)S\psi(1)\mbox{Tr}\left[SG(1-2)UG(2-1)\right]\psi^{\dagger}(2)U\psi(2),
\end{eqnarray}
where, in the spinless case,
$\mathcal{G}=\{A_1,A_2,D_1,D_2,C_3,D_3\}$ and the corresponding
couplings, in order of appearance of $S$ in $\mathcal{G}$, are
$\{g_{A_1},g_{A_2},g_{A_2},g_{D_2},g_{\gamma},g_{\gamma}\}$. Using
the gradient expansion to determine the RG fate of the marginal
couplings, we find that
\begin{eqnarray}
\Delta S^{(RPA)}_{eff}&=&\frac{1}{2}\sum_{S\in
\mathcal{G}}\sum_{U\in
\mathcal{G}}g_{S}g_{U}\int_1\psi^{\dagger}(1)S\psi(1)\mbox{Tr}
\left[-SU+\frac{1}{2}S\gamma_1U\gamma_1+\frac{1}{2}S\gamma_2U\gamma_2\right]
\psi^{\dagger}(1)U\psi(1)\frac{m}{4\pi}\ln s.
\end{eqnarray}
Performing the traces gives
\begin{eqnarray}
\Delta
S^{(RPA)}_{eff}=-\int_1\left(2g^2_{A_2}\left[(\psi^{\dagger}A_2\psi)^2+(\psi^{\dagger}D_1\psi)^2\right]+4g^2_{D_2}(\psi^{\dagger}D_2\psi)^2+4g^2_{\gamma}
\left[(\psi^{\dagger}C_3\psi)^2+(\psi^{\dagger}D_3\psi)^2\right]\right)\frac{m}{4\pi}\ln
s.
\end{eqnarray}
For the second and third (vertex) diagrams in Figure
\ref{fig:graphs} we have the following terms:
\begin{eqnarray}\label{eq:RGvertex}
\Delta
S^{(V)}_{eff}&=&-\sum_{S\in\mathcal{G}}\sum_{U\in\mathcal{G}}g_{S}g_{U}\int_1\int_2\psi^{\dagger}(1)S
G(1-2)UG(2-1)S\psi(1)\psi^{\dagger}(2)U\psi(2).
\end{eqnarray}
Performing the gradient expansion gives
\begin{eqnarray}
\Delta S^{(V)}_{eff}&=&-\sum_{S\in \mathcal{G}}\sum_{U\in
\mathcal{G}}g_{S}g_{U}\int_1\psi^{\dagger}(1)S\left(-U+\frac{1}{2}\gamma_1U\gamma_1+\frac{1}{2}\gamma_2U\gamma_2\right)S\psi(1)\psi^{\dagger}(1)U\psi(1)
\frac{m}{4\pi}\ln s.
\end{eqnarray}
Performing the requisite sums and matrix algebra gives
\begin{eqnarray}
\Delta
S^{(V)}_{eff}&=&-\int_1\left(g_{A_2}(-g_{A_1}+g_{D_2}+2g_{\gamma})\left[(\psi^{\dagger}A_2\psi)^2+(\psi^{\dagger}D_1\psi)^2\right]
-2g_{D_2}(g_{A_1}-2g_{A_2}+g_{D_2}+2g_{\gamma})(\psi^{\dagger}D_2\psi)^2\right)\frac{m}{4\pi}\ln
s\nonumber\\
&-&\int_1\left(-2g_{\gamma}(g_{A_1}-2g_{A_2}+g_{D_2})
\left[(\psi^{\dagger}C_3\psi)^2+(\psi^{\dagger}D_3\psi)^2\right]\right)\frac{m}{4\pi}\ln
s.
\end{eqnarray}
For the fourth and fifth diagrams in Figure \ref{fig:graphs} we find
the following terms:
\begin{eqnarray}\label{eq:RGladders}
\Delta
S^{(L)}_{eff}&=&-\frac{1}{4}\sum_{S,U\in\mathcal{G}}g_{S}g_{U}\int_1\left(\left(\psi^{\dagger}(1)\left[S,U\right]\psi(1)\right)^2
+\frac{1}{2}\sum_{a=1}^2\left(\psi^{\dagger}(1)\left(S\gamma_{a}U+U\gamma_{a}S\right)\psi(1)\right)^2\right)\frac{m}{4\pi}\ln
s.\end{eqnarray} The corresponding change in the effective action is
\begin{eqnarray}
\Delta
S^{(L)}_{eff}&=&-\int_1\left(2g_{A_1}g_{A_2}(\psi^{\dagger}A_1\psi)^2+
\frac{1}{2}(g^2_{A_1}+4g^2_{A_2}-4g_{A_2}g_{D_2}+2g^2_{D_2}+2g^2_{\gamma})\left[(\psi^{\dagger}A_2\psi)^2+(\psi^{\dagger}D_1\psi)^2\right]
\right)\nonumber\\
&-&\int_1\left(2g_{A_2}(g_{D_2}-g_{A_2})(\psi^{\dagger}D_2\psi)^2+
2g_{A_2}g_{\gamma}\left[(\psi^{\dagger}C_3\psi)^2+(\psi^{\dagger}D_3\psi)^2\right]-2g^2_{\gamma}(\psi^{\dagger}B_2\psi)^2\right)\nonumber\\
&-&\int_1\left((g_{D_2}-2g_{A_2})g_{\gamma}\left[(\psi^{\dagger}A_4\psi)^2+(\psi^{\dagger}B_4\psi)^2+(\psi^{\dagger}C_4\psi)^2+(\psi^{\dagger}D_4\psi)^2\right]\right).
\end{eqnarray}
In order to find the renormalization of the coupling constants, the
last two terms must be rewritten using the Fierz identities
(\ref{eq:FierzB2}) and (\ref{eq:FierzA4}). Adding the terms from
$\Delta \mathcal{S}_{eff}^{(RPA)}$, $\Delta
\mathcal{S}_{eff}^{(V)}$, and $\Delta \mathcal{S}_{eff}^{(L)}$,
rescaling the fields and the integration measure, and comparing to
the starting action (\ref{eq:LintAfterFierz}) we find the RG
equations (\ref{eq:RGspinless}) for the four coupling constants in
the spinless case.

The above equations (\ref{eq:RGrpa}), (\ref{eq:RGvertex}) and
(\ref{eq:RGladders}) can also be used to derive the flow equations
of the $9$ coupling constants in the case of spin-$1/2$ fermions. In
this case, we have $\mathcal{G}=\{A_1\otimes 1,A_2\otimes
1,D_1\otimes 1,B_1\otimes 1,C_2\otimes 1,B_2\otimes 1,C_1\otimes
1,D_2\otimes 1,A_3\otimes 1,B_3\otimes 1,B_4\otimes 1,A_4\otimes
1,C_4\otimes 1,D_4\otimes 1,C_3\otimes 1,D_3\otimes 1\}$ and the
corresponding couplings, in order of appearance of $S$ in
$\mathcal{G}$, are $\{g^{(c)}_{A_1}, g^{(c)}_{A_2}, g^{(c)}_{A_2},
g^{(c)}_{B_1}, g^{(c)}_{B_1}, g^{(c)}_{B_2}, g^{(c)}_{C_1},
g^{(c)}_{D_2},g^{(c)}_{\alpha},g^{(c)}_{\alpha},g^{(c)}_{\beta},g^{(c)}_{\beta},g^{(c)}_{\beta},g^{(c)}_{\beta},g^{(c)}_{\gamma},g^{(c)}_{\gamma}\}$.

The resulting RG flow equations for spin-$1/2$ fermions are
\begin{eqnarray}\label{eq:RGspinahalf}\label{eq:RGspinfulA1}
\frac{dg^{(c)}_{A_1}}{d\ln s}&=&-4\left(g^{(c)}_{A_1}g^{(c)}_{A_2}+g^{(c)}_{B_1}g^{(c)}_{B_2}+2g^{(c)}_{\alpha}g^{(c)}_{\beta}\right)\frac{m}{4\pi}\\
\frac{dg^{(c)}_{A_2}}{d\ln
s}&=&\left(-{g^{(c)}_{A_1}}^2+2g^{(c)}_{A_1}g^{(c)}_{A_2}
-12{g^{(c)}_{A_2}}^2-{g^{(c)}_{B_2}}^2-
(g^{(c)}_{C_1}-2g^{(c)}_{B_1})^2 -{g^{(c)}_{D_2}}^2
+2g^{(c)}_{A_2}\left(g^{(c)}_{B_2}-g^{(c)}_{C_1}+g^{(c)}_{D_2}+
2g^{(c)}_{\alpha}-2g^{(c)}_{\gamma}\right)\right.\nonumber\\
&-&\left.
2\left({g^{(c)}_{\alpha}}^2+\left(g^{(c)}_{\gamma}-2g^{(c)}_{\beta}\right)^2\right)
\right)\frac{m}{4\pi}\\
\frac{dg^{(c)}_{B_1}}{d\ln
s}&=&\left(2g^{(c)}_{B_1}\left(g^{(c)}_{A_1}-4
g^{(c)}_{B_1}-4g^{(c)}_{A_2}+g^{(c)}_{B_2}-g^{(c)}_{C_1}
+g^{(c)}_{D_2}-2 g^{(c)}_{\alpha}+2g^{(c)}_{\gamma}
\right)\right.\nonumber\\
&-&\left. 2\left(g^{(c)}_{A_1}
g^{(c)}_{B_2}-g^{(c)}_{C_1}(2g^{(c)}_{A_2}-
g^{(c)}_{D_2})-2g^{(c)}_{\alpha}
(2g^{(c)}_{\beta}-g^{(c)}_{\gamma})\right)\right)\frac{m}{4\pi}\nonumber\\
\frac{dg^{(c)}_{B_2}}{d\ln
s}&=&-4\left(g^{(c)}_{A_1}g^{(c)}_{B_1}+g^{(c)}_{A_2}g^{(c)}_{B_2}
-{g^{(c)}_{\alpha}}^2-2{g^{(c)}_{\beta}}^2+2g^{(c)}_{\beta}g^{(c)}_{\gamma}-
{g^{(c)}_{\gamma}}^2 \right)\frac{m}{4\pi}\\
\frac{dg^{(c)}_{C_1}}{d\ln
s}&=&4\left(g^{(c)}_{C_1}\left(g^{(c)}_{A_1}-3g^{(c)}_{A_2}-2g^{(c)}_{B_1}+g^{(c)}_{B_2}
-3g^{(c)}_{C_1}+g^{(c)}_{D_2}-2g^{(c)}_{\alpha}+4g^{(c)}_{\beta}-2g^{(c)}_{\gamma}
\right)\right.\nonumber\\
&+&\left.
g^{(c)}_{B_1}(2g^{(c)}_{A_2}-g^{(c)}_{D_2})-2g^{(c)}_{\alpha}(g^{(c)}_{\beta}-g^{(c)}_{\gamma})
\right)\frac{m}{4\pi}\\
\frac{dg^{(c)}_{D_2}}{d\ln s}&=& 4\left( g^{(c)}_{D_2}\left(
g^{(c)}_{A_1}-3g^{(c)}_{A_2}-2g^{(c)}_{B_1}+g^{(c)}_{B_2}+g^{(c)}_{C_1}-3g^{(c)}_{D_2}
+2g^{(c)}_{\alpha}-4g^{(c)}_{\beta}+2g^{(c)}_{\gamma}
\right)\right.\nonumber\\
&+&\left.
{g^{(c)}_{A_2}}^2+g^{(c)}_{B_1}\left(g^{(c)}_{B_1}-g^{(c)}_{C_1}\right)+
2g^{(c)}_{\beta}\left(g^{(c)}_{\beta}-g^{(c)}_{\gamma}\right)\right)\frac{m}{4\pi}\\
\frac{dg^{(c)}_{\alpha}}{d\ln
s}&=&-4\left(g^{(c)}_{\alpha}\left(g^{(c)}_{A_2}-g^{(c)}_{B_2}\right)+
g^{(c)}_{\beta}\left(g^{(c)}_{A_1}-2g^{(c)}_{B_1}+g^{(c)}_{C_1}\right)+
g^{(c)}_{\gamma}\left(g^{(c)}_{B_1}-g^{(c)}_{C_1}\right)\right)\frac{m}{4\pi}\\
\frac{dg^{(c)}_{\beta}}{d\ln
s}&=&2\left(g^{(c)}_{\beta}\left(g^{(c)}_{A_1}-4g^{(c)}_{A_2}+g^{(c)}_{B_2}+
g^{(c)}_{C_1}+g^{(c)}_{D_2}-4g^{(c)}_{\beta}
\right)-g^{(c)}_{\alpha}\left(g^{(c)}_{A_1}-2g^{(c)}_{B_1}+g^{(c)}_{C_1}\right)
+g^{(c)}_{\gamma}\left(2g^{(c)}_{A_2}-g^{(c)}_{B_2}-g^{(c)}_{D_2}\right)
\right)\frac{m}{4\pi}\nonumber\\\\
\label{eq:RGspinfulgamma} \frac{dg^{(c)}_{\gamma}}{d\ln
s}&=&-4\left(g^{(c)}_{\alpha}\left(g^{(c)}_{B_1}-g^{(c)}_{C_1}\right)
+g^{(c)}_{\beta}\left(g^{(c)}_{B_2}-2g^{(c)}_{A_2}+g^{(c)}_{D_2}\right)
-g^{(c)}_{\gamma}\left(g^{(c)}_{A_1}-3g^{(c)}_{A_2}+2g^{(c)}_{B_1}
-g^{(c)}_{C_1}+g^{(c)}_{D_2}-4g^{(c)}_{\gamma}\right)
\right)\frac{m}{4\pi}.
\end{eqnarray}
\section{Fierz reduction for spin-$\frac{1}{2}$ case}
If we arrange the quartic contact terms into an 18-dimensional
vector
\begin{eqnarray}\label{eq:FierzVectorSpinHalf}
\mathcal{V}&=&\left\{(\psi^{\dagger}_{\alpha}A_1\psi_{\alpha})^2,(\psi^{\dagger}_{\alpha}A_2\psi_{\alpha})^2+(\psi^{\dagger}_{\alpha}D_1\psi_{\alpha})^2,
(\psi^{\dagger}_{\alpha}B_1\psi_{\alpha})^2+(\psi^{\dagger}_{\alpha}C_2\psi_{\alpha})^2,
(\psi^{\dagger}_{\alpha}B_2\psi_{\alpha})^2,(\psi^{\dagger}_{\alpha}C_1\psi_{\alpha})^2,\right.\nonumber\\
&&\left.
(\psi^{\dagger}_{\alpha}D_2\psi_{\alpha})^2,(\psi^{\dagger}_{\alpha}A_3\psi_{\alpha})^2+(\psi^{\dagger}_{\alpha}B_3\psi_{\alpha})^2,
\sum_{X=A,B,C,D}(\psi^{\dagger}_{\alpha}X_4\psi_{\alpha})^2,
(\psi^{\dagger}_{\alpha}C_3\psi_{\alpha})^2+(\psi^{\dagger}_{\alpha}D_3\psi_{\alpha})^2,\right.\nonumber\\
&&\left.
(\psi^{\dagger}_{\alpha}A_1\vec{\sigma}_{\alpha\beta}\psi_{\beta})^2,(\psi^{\dagger}_{\alpha}A_2\vec{\sigma}_{\alpha\beta}\psi_{\beta})^2+
(\psi^{\dagger}_{\alpha}D_1\vec{\sigma}_{\alpha\beta}\psi_{\beta})^2,
(\psi^{\dagger}_{\alpha}B_1\vec{\sigma}_{\alpha\beta}\psi_{\beta})^2+
(\psi^{\dagger}_{\alpha}C_2\vec{\sigma}_{\alpha\beta}\psi_{\beta})^2,
(\psi^{\dagger}_{\alpha}B_2\vec{\sigma}_{\alpha\beta}\psi_{\beta})^2,
(\psi^{\dagger}_{\alpha}C_1\vec{\sigma}_{\alpha\beta}\psi_{\beta})^2,\right.\nonumber\\
&&\left.(\psi^{\dagger}_{\alpha}D_2\vec{\sigma}_{\alpha\beta}\psi_{\beta})^2,
(\psi^{\dagger}_{\alpha}A_3\vec{\sigma}_{\alpha\beta}\psi_{\beta})^2+(\psi^{\dagger}_{\alpha}B_3\vec{\sigma}_{\alpha\beta}\psi_{\beta})^2,
\sum_{X=A,B,C,D}(\psi^{\dagger}_{\alpha}X_4\vec{\sigma}_{\alpha\beta}\psi_{\beta})^2,
(\psi^{\dagger}_{\alpha}C_3\vec{\sigma}_{\alpha\beta}\psi_{\beta})^2+(\psi^{\dagger}_{\alpha}D_3\vec{\sigma}_{\alpha\beta}\psi_{\beta})^2,\right\}\nonumber\\
\end{eqnarray}
then the Fierz identities (\ref{eq:Fierz8by8}) can be used to relate
different components of $\mathcal{V}$ via the linear constraint
$\mathcal{F}\mathcal{V}=0$.

In practice, to obtain the Fierz matrix $\mathcal{F}$ we arrange the
64 $SU(8)$ generators $\Lambda^a$ in the order $\{X\otimes 1,
X\otimes \sigma^z, X\otimes \sigma^x, X\otimes \sigma^y\}$, where
$X=\{A_1,A_2,D_1,B_1,C_2,B_2,C_1,D_2,A_3,B_3,B_4,A_4,C_4,D_4,C_3,D_3\}$.
Straightforwardly, a 64-component analog of the Fierz vector,
obtained by using our ordered set of $SU(8)$ generators, will be
denoted by $\tilde{\mathcal{V}}$. Next, we notice that only the
(diagonal) terms with $S=T$ enter our $\mathcal{L}_{int}$, and that
$\mbox{Tr}[S\Lambda^a S\Lambda^b]\sim \delta_{ab}$. Therefore, for
$\Lambda^a$ arranged as described above, we numerically generate a
64$\times$64 matrix $\tilde{\Phi}_{ab}=-\mbox{Tr}[\Lambda^a\Lambda^b
\Lambda^a\Lambda^b]/64$. We then construct an auxiliary 64$\times$64
matrix
$$
\mathcal{M}=\left(\begin{array}{cccc} \mathcal{Q} & 0 & 0 & 0\\
0 & \mathcal{Q} & \mathcal{Q} & \mathcal{Q} \\
0 & 2\mathcal{Q} & -\mathcal{Q} & -\mathcal{Q} \\
0 & 0 & \mathcal{Q} & -\mathcal{Q} \\
\end{array}\right)
$$
where
$$
\mathcal{Q}=\left(
\begin{array}{cccccccccccccccc}
1& 0& 0& 0& 0& 0& 0& 0& 0& 0& 0& 0& 0& 0& 0& 0\\ 0& 1& 1& 0& 0& 0&0&
0& 0& 0& 0& 0& 0& 0& 0& 0\\ 0& 0& 0& 1& 1& 0& 0& 0& 0& 0& 0&
0& 0& 0& 0& 0\\ 0& 0& 0& 0& 0& 1& 0& 0& 0& 0& 0& 0& 0& 0& 0& 0\\
0& 0& 0& 0& 0& 0& 1& 0& 0& 0& 0& 0& 0& 0& 0& 0\\ 0& 0& 0& 0& 0& 0&
0& 1& 0& 0& 0& 0& 0& 0& 0& 0\\ 0& 0& 0& 0& 0& 0& 0& 0& 1& 1& 0& 0&
0& 0& 0& 0\\ 0& 0& 0& 0& 0& 0& 0& 0& 0& 0& 1& 1& 1& 1& 0& 0\\ 0& 0&
0& 0& 0& 0& 0& 0& 0& 0& 0& 0& 0& 0& 1& 1\\ 0& -1& 1& 0& 0& 0& 0& 0&
0& 0& 0& 0& 0& 0& 0& 0\\ 0& 0& 0& -1& 1& 0& 0& 0& 0& 0& 0& 0& 0& 0&
0& 0\\ 0& 0& 0& 0& 0& 0& 0& 0& -1& 1& 0& 0& 0& 0& 0& 0\\ 0& 0& 0& 0&
0& 0& 0& 0& 0& 0& -1& 1& -1& 1& 0& 0\\ 0& 0& 0& 0& 0& 0& 0& 0& 0& 0&
-1& -1& 1& 1& 0& 0\\ 0& 0& 0& 0& 0& 0& 0& 0& 0& 0& 1& -1& -1& 1& 0&
0\\ 0& 0& 0& 0& 0& 0& 0& 0& 0& 0& 0& 0& 0& 0& -1& 1
\end{array}\right)
$$
in order to write
$$
\tilde{\mathcal{V}}=\tilde{\Phi}
\tilde{\mathcal{V}}\;\;\;\Rightarrow\;\;\:
\mathcal{M}\tilde{\mathcal{V}}=\left(\mathcal{M}\tilde{\Phi}
\mathcal{M}^{-1}\right)\mathcal{M} \tilde{\mathcal{V}}.
$$
It is easily seen that components $1-9$ and $17-25$ of
$\mathcal{M}\tilde{\mathcal{V}}$ correspond to our Fierz vector
$\mathcal{V}$. Moreover, the blocks $1-9$ and $17-25$ of the matrix
$\mathcal{M}\tilde{\Phi} \mathcal{M}^{-1}$ do not couple to the rest
of the components, and, when subtracted from an 18-dimensional unit
matrix, correspond to the sought Fierz matrix,
\begin{eqnarray}
\label{eq:FierzMatrixSpinHalf}
 \mathcal{F}&=&\left(
\begin{array}{cccccccccccccccccc}
9& 1& 1& 1& 1& 1& 1& 1& 1& 1& 1& 1& 1& 1& 1& 1& 1& 1\\
2& 8& 0& 2& -2& -2& 2& 0& -2& 2& 0& 0& 2& -2& -2& 2& 0& -2\\
2& 0& 8& 2& -2& -2& -2& 0& 2& 2& 0& 0& 2& -2& -2& -2& 0& 2\\
1& 1& 1& 9& 1& 1& -1& -1& -1& 1& 1& 1& 1& 1& 1& -1& -1& -1\\
1& -1& -1& 1& 9& 1& -1& 1& -1& 1& -1& -1& 1& 1& 1& -1& 1& -1\\
1& -1& -1& 1& 1& 9& 1& -1& 1& 1& -1& -1& 1& 1& 1& 1& -1& 1\\
2& 2& -2& -2& -2& 2& 8& 0& 0& 2& 2& -2& -2& -2& 2& 0& 0& 0\\
4& 0& 0& -4& 4& -4& 0& 8& 0& 4& 0& 0& -4& 4& -4& 0& 0& 0\\
2& -2& 2& -2& -2& 2& 0& 0& 8& 2& -2& 2& -2& -2& 2& 0& 0& 0\\
3& 3& 3& 3& 3& 3& 3& 3& 3& 7& -1& -1& -1& -1& -1& -1& -1& -1\\
6& 0& 0& 6& -6& -6& 6& 0& -6& -2& 8& 0& -2& 2& 2& -2& 0& 2\\
6& 0& 0& 6& -6& -6& -6& 0& 6& -2& 0& 8& -2& 2& 2& 2& 0& -2\\
3& 3& 3& 3& 3& 3& -3& -3& -3& -1& -1& -1& 7& -1& -1& 1& 1& 1\\
3& -3& -3& 3& 3& 3& -3& 3& -3& -1& 1& 1& -1& 7& -1& 1& -1& 1\\
3&-3& -3& 3& 3& 3& 3& -3& 3& -1& 1& 1& -1& -1& 7& -1& 1& -1\\
6& 6& -6& -6& -6& 6& 0& 0& 0& -2& -2& 2& 2& 2& -2& 8& 0& 0\\
12& 0& 0& -12& 12& -12& 0& 0& 0& -4& 0& 0& 4& -4& 4& 0& 8& 0\\
6& -6& 6& -6& -6& 6& 0& 0& 0& -2& 2& -2& 2& 2& -2& 0& 0& 8
\end{array}
\right).
\end{eqnarray}
One can check that the above matrix has 9 zero eigenvalues, which
implies 9 independent couplings and 9 constraints. In addition, one
can solve the linear system $\mathcal{F}\mathcal{V}=0$ and eliminate
all terms of the form
$(\psi^{\dagger}_{\alpha}X_j\vec{\sigma}_{\alpha\beta}\psi_{\beta})^2$
in favor of $(\psi^{\dagger}_{\alpha}X_j\psi_{\alpha})^2$.

\end{widetext}

\bibliography{rg0810}

\begin{thebibliography}{26}
\expandafter\ifx\csname natexlab\endcsname\relax\def\natexlab#1{#1}\fi
\expandafter\ifx\csname bibnamefont\endcsname\relax
  \def\bibnamefont#1{#1}\fi
\expandafter\ifx\csname bibfnamefont\endcsname\relax
  \def\bibfnamefont#1{#1}\fi
\expandafter\ifx\csname citenamefont\endcsname\relax
  \def\citenamefont#1{#1}\fi
\expandafter\ifx\csname url\endcsname\relax
  \def\url#1{\texttt{#1}}\fi
\expandafter\ifx\csname urlprefix\endcsname\relax\def\urlprefix{URL }\fi
\providecommand{\bibinfo}[2]{#2}
\providecommand{\eprint}[2][]{\url{#2}}

\bibitem[{\citenamefont{McCann and
  Fal'ko}(2006)}]{McCannFalko2006PhysRevLett.96.086805}
\bibinfo{author}{\bibfnamefont{E.}~\bibnamefont{McCann}} \bibnamefont{and}
  \bibinfo{author}{\bibfnamefont{V.~I.} \bibnamefont{Fal'ko}},
  \bibinfo{journal}{Phys. Rev. Lett.} \textbf{\bibinfo{volume}{96}},
  \bibinfo{pages}{086805} (\bibinfo{year}{2006}).

\bibitem[{\citenamefont{Castro~Neto et~al.}(2009)\citenamefont{Castro~Neto,
  Guinea, Peres, Novoselov, and Geim}}]{CastroNeto2009RevModPhys.81.109}
\bibinfo{author}{\bibfnamefont{A.~H.} \bibnamefont{Castro~Neto}},
  \bibinfo{author}{\bibfnamefont{F.}~\bibnamefont{Guinea}},
  \bibinfo{author}{\bibfnamefont{N.~M.~R.} \bibnamefont{Peres}},
  \bibinfo{author}{\bibfnamefont{K.~S.} \bibnamefont{Novoselov}},
  \bibnamefont{and} \bibinfo{author}{\bibfnamefont{A.~K.} \bibnamefont{Geim}},
  \bibinfo{journal}{Rev. Mod. Phys.} \textbf{\bibinfo{volume}{81}},
  \bibinfo{pages}{109} (\bibinfo{year}{2009}).

\bibitem[{\citenamefont{Vafek and
  Yang}(2010)}]{VafekYang2010PhysRevB.81.041401}
\bibinfo{author}{\bibfnamefont{O.}~\bibnamefont{Vafek}} \bibnamefont{and}
  \bibinfo{author}{\bibfnamefont{K.}~\bibnamefont{Yang}},
  \bibinfo{journal}{Phys. Rev. B} \textbf{\bibinfo{volume}{81}},
  \bibinfo{pages}{041401} (\bibinfo{year}{2010}).

\bibitem[{\citenamefont{Zhang et~al.}(2010)\citenamefont{Zhang, Min, Polini,
  and MacDonald}}]{FanZhangPRB2010PhysRevB.81.041402}
\bibinfo{author}{\bibfnamefont{F.}~\bibnamefont{Zhang}},
  \bibinfo{author}{\bibfnamefont{H.}~\bibnamefont{Min}},
  \bibinfo{author}{\bibfnamefont{M.}~\bibnamefont{Polini}}, \bibnamefont{and}
  \bibinfo{author}{\bibfnamefont{A.~H.} \bibnamefont{MacDonald}},
  \bibinfo{journal}{Phys. Rev. B} \textbf{\bibinfo{volume}{81}},
  \bibinfo{pages}{041402} (\bibinfo{year}{2010}).

\bibitem[{\citenamefont{Nandkishore and
  Levitov}(2010)}]{LevitovPRL2010PhysRevLett.104.156803}
\bibinfo{author}{\bibfnamefont{R.}~\bibnamefont{Nandkishore}} \bibnamefont{and}
  \bibinfo{author}{\bibfnamefont{L.}~\bibnamefont{Levitov}},
  \bibinfo{journal}{Phys. Rev. Lett.} \textbf{\bibinfo{volume}{104}},
  \bibinfo{pages}{156803} (\bibinfo{year}{2010}).

\bibitem[{\citenamefont{{Lemonik} et~al.}(2010)\citenamefont{{Lemonik},
  {Aleiner}, {Toke}, and {Fal'ko}}}]{Lemonik2010arXiv1006.1399L}
\bibinfo{author}{\bibfnamefont{Y.}~\bibnamefont{{Lemonik}}},
  \bibinfo{author}{\bibfnamefont{I.~L.} \bibnamefont{{Aleiner}}},
  \bibinfo{author}{\bibfnamefont{C.}~\bibnamefont{{Toke}}}, \bibnamefont{and}
  \bibinfo{author}{\bibfnamefont{V.~I.} \bibnamefont{{Fal'ko}}},
  \bibinfo{journal}{ArXiv e-prints}  (\bibinfo{year}{2010}),
  \eprint{1006.1399}.

\bibitem[{\citenamefont{Min et~al.}(2008)\citenamefont{Min, Borghi, Polini, and
  MacDonald}}]{Min2008PhysRevB.77.041407}
\bibinfo{author}{\bibfnamefont{H.}~\bibnamefont{Min}},
  \bibinfo{author}{\bibfnamefont{G.}~\bibnamefont{Borghi}},
  \bibinfo{author}{\bibfnamefont{M.}~\bibnamefont{Polini}}, \bibnamefont{and}
  \bibinfo{author}{\bibfnamefont{A.~H.} \bibnamefont{MacDonald}},
  \bibinfo{journal}{Phys. Rev. B} \textbf{\bibinfo{volume}{77}},
  \bibinfo{pages}{041407} (\bibinfo{year}{2008}).

\bibitem[{\citenamefont{Sun et~al.}(2009)\citenamefont{Sun, Yao, Fradkin, and
  Kivelson}}]{KaiSun2009PhysRevLett.103.046811}
\bibinfo{author}{\bibfnamefont{K.}~\bibnamefont{Sun}},
  \bibinfo{author}{\bibfnamefont{H.}~\bibnamefont{Yao}},
  \bibinfo{author}{\bibfnamefont{E.}~\bibnamefont{Fradkin}}, \bibnamefont{and}
  \bibinfo{author}{\bibfnamefont{S.~A.} \bibnamefont{Kivelson}},
  \bibinfo{journal}{Phys. Rev. Lett.} \textbf{\bibinfo{volume}{103}},
  \bibinfo{pages}{046811} (\bibinfo{year}{2009}).

\bibitem[{\citenamefont{Raghu et~al.}(2010)\citenamefont{Raghu, Kivelson, and
  Scalapino}}]{RaghuKivelsonScalapino2010PhysRevB.81.224505}
\bibinfo{author}{\bibfnamefont{S.}~\bibnamefont{Raghu}},
  \bibinfo{author}{\bibfnamefont{S.~A.} \bibnamefont{Kivelson}},
  \bibnamefont{and} \bibinfo{author}{\bibfnamefont{D.~J.}
  \bibnamefont{Scalapino}}, \bibinfo{journal}{Phys. Rev. B}
  \textbf{\bibinfo{volume}{81}}, \bibinfo{pages}{224505}
  (\bibinfo{year}{2010}).

\bibitem[{\citenamefont{{Novoselov} et~al.}(2006)\citenamefont{{Novoselov},
  {McCann}, {Morozov}, {Fal'Ko}, {Katsnelson}, {Zeitler}, {Jiang}, {Schedin},
  and {Geim}}}]{Geim2006NatPhys}
\bibinfo{author}{\bibfnamefont{K.~S.} \bibnamefont{{Novoselov}}},
  \bibinfo{author}{\bibfnamefont{E.}~\bibnamefont{{McCann}}},
  \bibinfo{author}{\bibfnamefont{S.~V.} \bibnamefont{{Morozov}}},
  \bibinfo{author}{\bibfnamefont{V.~I.} \bibnamefont{{Fal'Ko}}},
  \bibinfo{author}{\bibfnamefont{M.~I.} \bibnamefont{{Katsnelson}}},
  \bibinfo{author}{\bibfnamefont{U.}~\bibnamefont{{Zeitler}}},
  \bibinfo{author}{\bibfnamefont{D.}~\bibnamefont{{Jiang}}},
  \bibinfo{author}{\bibfnamefont{F.}~\bibnamefont{{Schedin}}},
  \bibnamefont{and} \bibinfo{author}{\bibfnamefont{A.~K.}
  \bibnamefont{{Geim}}}, \bibinfo{journal}{Nature Physics}
  \textbf{\bibinfo{volume}{2}}, \bibinfo{pages}{177} (\bibinfo{year}{2006}).

\bibitem[{\citenamefont{{Feldman} et~al.}(2009)\citenamefont{{Feldman},
  {Martin}, and {Yacoby}}}]{Yacoby2009NatPhys889}
\bibinfo{author}{\bibfnamefont{B.~E.} \bibnamefont{{Feldman}}},
  \bibinfo{author}{\bibfnamefont{J.}~\bibnamefont{{Martin}}}, \bibnamefont{and}
  \bibinfo{author}{\bibfnamefont{A.}~\bibnamefont{{Yacoby}}},
  \bibinfo{journal}{Nature Physics} \textbf{\bibinfo{volume}{5}},
  \bibinfo{pages}{889} (\bibinfo{year}{2009}), \eprint{0909.2883}.

\bibitem[{\citenamefont{Zhao et~al.}(2010)\citenamefont{Zhao, Cadden-Zimansky,
  Jiang, and Kim}}]{PhilipKim2010PhysRevLett.104.066801}
\bibinfo{author}{\bibfnamefont{Y.}~\bibnamefont{Zhao}},
  \bibinfo{author}{\bibfnamefont{P.}~\bibnamefont{Cadden-Zimansky}},
  \bibinfo{author}{\bibfnamefont{Z.}~\bibnamefont{Jiang}}, \bibnamefont{and}
  \bibinfo{author}{\bibfnamefont{P.}~\bibnamefont{Kim}},
  \bibinfo{journal}{Phys. Rev. Lett.} \textbf{\bibinfo{volume}{104}},
  \bibinfo{pages}{066801} (\bibinfo{year}{2010}).

\bibitem[{\citenamefont{{Ohta} et~al.}(2006)\citenamefont{{Ohta}, {Bostwick},
  {Seyller}, {Horn}, and {Rotenberg}}}]{RotenbergARPES2006Science313}
\bibinfo{author}{\bibfnamefont{T.}~\bibnamefont{{Ohta}}},
  \bibinfo{author}{\bibfnamefont{A.}~\bibnamefont{{Bostwick}}},
  \bibinfo{author}{\bibfnamefont{T.}~\bibnamefont{{Seyller}}},
  \bibinfo{author}{\bibfnamefont{K.}~\bibnamefont{{Horn}}}, \bibnamefont{and}
  \bibinfo{author}{\bibfnamefont{E.}~\bibnamefont{{Rotenberg}}},
  \bibinfo{journal}{Science} \textbf{\bibinfo{volume}{313}},
  \bibinfo{pages}{951} (\bibinfo{year}{2006}).

\bibitem[{\citenamefont{Yan et~al.}(2008)\citenamefont{Yan, Henriksen, Kim, and
  Pinczuk}}]{Pinczuk2008PhysRevLett.101.136804}
\bibinfo{author}{\bibfnamefont{J.}~\bibnamefont{Yan}},
  \bibinfo{author}{\bibfnamefont{E.~A.} \bibnamefont{Henriksen}},
  \bibinfo{author}{\bibfnamefont{P.}~\bibnamefont{Kim}}, \bibnamefont{and}
  \bibinfo{author}{\bibfnamefont{A.}~\bibnamefont{Pinczuk}},
  \bibinfo{journal}{Phys. Rev. Lett.} \textbf{\bibinfo{volume}{101}},
  \bibinfo{pages}{136804} (\bibinfo{year}{2008}).

\bibitem[{\citenamefont{{Zhang} et~al.}(2009)\citenamefont{{Zhang}, {Tang},
  {Girit}, {Hao}, {Martin}, {Zettl}, {Crommie}, {Shen}, and
  {Wang}}}]{FengWangIR2009Natur.459..820Z}
\bibinfo{author}{\bibfnamefont{Y.}~\bibnamefont{{Zhang}}},
  \bibinfo{author}{\bibfnamefont{T.}~\bibnamefont{{Tang}}},
  \bibinfo{author}{\bibfnamefont{C.}~\bibnamefont{{Girit}}},
  \bibinfo{author}{\bibfnamefont{Z.}~\bibnamefont{{Hao}}},
  \bibinfo{author}{\bibfnamefont{M.~C.} \bibnamefont{{Martin}}},
  \bibinfo{author}{\bibfnamefont{A.}~\bibnamefont{{Zettl}}},
  \bibinfo{author}{\bibfnamefont{M.~F.} \bibnamefont{{Crommie}}},
  \bibinfo{author}{\bibfnamefont{Y.~R.} \bibnamefont{{Shen}}},
  \bibnamefont{and} \bibinfo{author}{\bibfnamefont{F.}~\bibnamefont{{Wang}}},
  \bibinfo{journal}{\nat} \textbf{\bibinfo{volume}{459}}, \bibinfo{pages}{820}
  (\bibinfo{year}{2009}).

\bibitem[{\citenamefont{{Nandkishore} and
  {Levitov}}(2010)}]{NandkishoreQAH2010arXiv1002.1966N}
\bibinfo{author}{\bibfnamefont{R.}~\bibnamefont{{Nandkishore}}}
  \bibnamefont{and}
  \bibinfo{author}{\bibfnamefont{L.}~\bibnamefont{{Levitov}}},
  \bibinfo{journal}{ArXiv e-prints}  (\bibinfo{year}{2010}),
  \eprint{1002.1966}.

\bibitem[{\citenamefont{Herbut et~al.}(2009)\citenamefont{Herbut, Juri\ifmmode
  \check{c}\else \v{c}\fi{}i\ifmmode~\acute{c}\else \'{c}\fi{}, and
  Roy}}]{HerbutJuricicRoy2009PhysRevB.79.085116}
\bibinfo{author}{\bibfnamefont{I.~F.} \bibnamefont{Herbut}},
  \bibinfo{author}{\bibfnamefont{V.}~\bibnamefont{Juri\ifmmode \check{c}\else
  \v{c}\fi{}i\ifmmode~\acute{c}\else \'{c}\fi{}}}, \bibnamefont{and}
  \bibinfo{author}{\bibfnamefont{B.}~\bibnamefont{Roy}},
  \bibinfo{journal}{Phys. Rev. B} \textbf{\bibinfo{volume}{79}},
  \bibinfo{pages}{085116} (\bibinfo{year}{2009}).

\bibitem[{\citenamefont{Bir and Pikus}(1974)}]{BirPikusBook1974}
\bibinfo{author}{\bibfnamefont{G.~L.} \bibnamefont{Bir}} \bibnamefont{and}
  \bibinfo{author}{\bibfnamefont{G.~E.} \bibnamefont{Pikus}},
  \emph{\bibinfo{title}{Symmetry and strain-indused effects in semiconductors}}
  (\bibinfo{publisher}{John Wiley}, \bibinfo{address}{New York},
  \bibinfo{year}{1974}).

\bibitem[{\citenamefont{Aleiner et~al.}(2007)\citenamefont{Aleiner, Kharzeev,
  and Tsvelik}}]{AleinerKharzeevTsvelik2007PhysRevB.76.195415}
\bibinfo{author}{\bibfnamefont{I.~L.} \bibnamefont{Aleiner}},
  \bibinfo{author}{\bibfnamefont{D.~E.} \bibnamefont{Kharzeev}},
  \bibnamefont{and} \bibinfo{author}{\bibfnamefont{A.~M.}
  \bibnamefont{Tsvelik}}, \bibinfo{journal}{Phys. Rev. B}
  \textbf{\bibinfo{volume}{76}}, \bibinfo{pages}{195415}
  (\bibinfo{year}{2007}).

\bibitem[{\citenamefont{Zhang et~al.}(2008)\citenamefont{Zhang, Li, Basov,
  Fogler, Hao, and Martin}}]{ZhangBasovFogler2008PhysRevB.78.235408}
\bibinfo{author}{\bibfnamefont{L.~M.} \bibnamefont{Zhang}},
  \bibinfo{author}{\bibfnamefont{Z.~Q.} \bibnamefont{Li}},
  \bibinfo{author}{\bibfnamefont{D.~N.} \bibnamefont{Basov}},
  \bibinfo{author}{\bibfnamefont{M.~M.} \bibnamefont{Fogler}},
  \bibinfo{author}{\bibfnamefont{Z.}~\bibnamefont{Hao}}, \bibnamefont{and}
  \bibinfo{author}{\bibfnamefont{M.~C.} \bibnamefont{Martin}},
  \bibinfo{journal}{Phys. Rev. B} \textbf{\bibinfo{volume}{78}},
  \bibinfo{pages}{235408} (\bibinfo{year}{2008}).

\bibitem[{\citenamefont{Nilsson et~al.}(2008)\citenamefont{Nilsson, Neto,
  Guinea, and Peres}}]{nilsson:045405}
\bibinfo{author}{\bibfnamefont{J.}~\bibnamefont{Nilsson}},
  \bibinfo{author}{\bibfnamefont{A.~H.~C.} \bibnamefont{Neto}},
  \bibinfo{author}{\bibfnamefont{F.}~\bibnamefont{Guinea}}, \bibnamefont{and}
  \bibinfo{author}{\bibfnamefont{N.~M.~R.} \bibnamefont{Peres}},
  \bibinfo{journal}{Phys. Rev. B} \textbf{\bibinfo{volume}{78}},
  \bibinfo{eid}{045405} (pages~\bibinfo{numpages}{34}) (\bibinfo{year}{2008}).

\bibitem[{\citenamefont{Shankar}(1994)}]{Shankar.RevModPhys.66.129.1994}
\bibinfo{author}{\bibfnamefont{R.}~\bibnamefont{Shankar}},
  \bibinfo{journal}{Rev. Mod. Phys.} \textbf{\bibinfo{volume}{66}},
  \bibinfo{pages}{129} (\bibinfo{year}{1994}).

\bibitem[{\citenamefont{Itzykson and Zuber}(2005)}]{ItzyksonZuberBook}
\bibinfo{author}{\bibfnamefont{C.}~\bibnamefont{Itzykson}} \bibnamefont{and}
  \bibinfo{author}{\bibfnamefont{J.-B.} \bibnamefont{Zuber}},
  \emph{\bibinfo{title}{Quantum Field Theory}} (\bibinfo{publisher}{Dover},
  \bibinfo{address}{Mineola, NY}, \bibinfo{year}{2005}), \bibinfo{note}{p.161}.

\bibitem[{\citenamefont{Haldane}(1988)}]{HaldanePRL1988PhysRevLett.61.2015}
\bibinfo{author}{\bibfnamefont{F.~D.~M.} \bibnamefont{Haldane}},
  \bibinfo{journal}{Phys. Rev. Lett.} \textbf{\bibinfo{volume}{61}},
  \bibinfo{pages}{2015} (\bibinfo{year}{1988}).

\bibitem[{\citenamefont{Fradkin}(1991)}]{FradkinBook}
\bibinfo{author}{\bibfnamefont{E.}~\bibnamefont{Fradkin}},
  \emph{\bibinfo{title}{Field Theories of Condensed Matter Systems}}
  (\bibinfo{publisher}{Addison-Wesley}, \bibinfo{address}{Redwood City, CA},
  \bibinfo{year}{1991}), \bibinfo{note}{ch. 2}.

\bibitem[{\citenamefont{Paiva et~al.}(2005)\citenamefont{Paiva, Scalettar,
  Zheng, Singh, and Oitmaa}}]{Paiva2005PhysRevB.72.085123}
\bibinfo{author}{\bibfnamefont{T.}~\bibnamefont{Paiva}},
  \bibinfo{author}{\bibfnamefont{R.~T.} \bibnamefont{Scalettar}},
  \bibinfo{author}{\bibfnamefont{W.}~\bibnamefont{Zheng}},
  \bibinfo{author}{\bibfnamefont{R.~R.~P.} \bibnamefont{Singh}},
  \bibnamefont{and} \bibinfo{author}{\bibfnamefont{J.}~\bibnamefont{Oitmaa}},
  \bibinfo{journal}{Phys. Rev. B} \textbf{\bibinfo{volume}{72}},
  \bibinfo{pages}{085123} (\bibinfo{year}{2005}).

\end{thebibliography}

\end{document}